\global\def\draftcontrol{0}

   \def\versionno{k-strings}

\catcode`\@=11

\expandafter\ifx\csname draftcontrol\endcsname\relax\global\def\draftcontrol{0}
\fi

{\count255=\time\divide\count255 by 60
\xdef\hourmin{\number\count255}
\multiply\count255 by-60\advance\count255 by\time
\xdef\hourmin{\hourmin:\ifnum\count255<10 0\fi\the\count255}}
\def\draftdate{\number\month/\number\day/\number\year\ \ \ \hourmin }

\newcommand\makepapertitle{\par
  \begingroup
    \renewcommand\thefootnote{\@fnsymbol\c@footnote}%
    \def\@makefnmark{\rlap{\@textsuperscript{\normalfont\@thefnmark}}}%
    \long\def\@makefntext##1{\parindent 1em\noindent
            \hb@xt@1.8em{%
                \hss\@textsuperscript{\normalfont\@thefnmark}}##1}%
     \newpage
     \global\@topnum\z@   
     \@makepapertitle
     \thispagestyle{empty}\@thanks
  \endgroup
  \setcounter{footnote}{0}%
  \global\let\thanks\relax
  \global\let\makepapertitle\relax
  \global\let\@makepapertitle\relax
  \global\let\@thanks\@empty
  \global\let\@author\@empty
  \global\let\@date\@empty
  \global\let\@title\@empty
  \global\let\title\relax
  \global\let\author\relax
  \global\let\date\relax
  \global\let\and\relax
  \def\version{\let\version\@version\@gobble}
}
\def\@makepapertitle{%
  \newpage
   \ifnum\draftcontrol=1 {}
   \version\versionno
   \vskip 3em%
   \else
   \hfill\hbox to 3cm {\parbox{4cm}{\@pubnum}\hss}%
   \vskip 3em%
   \fi
   \begin{center}%
   \let \footnote \thanks
     {\LARGE {\@title}}%
     \vskip 1.5em%
     {\normalsize
       \lineskip .5em%
       \begin{tabular}[t]{c}%
         \@author
       \end{tabular}\par}%
     \vskip 1.5em%
     {\@bstract}%
     \end{center}%
     \vskip 1.5em
     \@date%
   \par
}

\gdef\@pubnum{}
\def\pubnum#1{%
  \gdef\@pubnum{#1}}

\gdef\@bstract{}
\def\Abstract#1{%
  \gdef\@bstract{%
   \parbox{\textwidth-0pc}{%
   \centerline{\bf Abstract}\penalty1000%
\noindent
\renewcommand\baselinestretch{1.0}%
{#1}}}
}

\def\ps@paper{\let\@mkboth\@gobbletwo%
     \ifnum\draftcontrol=1
        \def\@oddfoot{\hbox to \textwidth{\tiny \versionno \hfil\tiny\draftdate}%
        \hskip -\textwidth \hbox to \textwidth{\hfil\rm\thepage\hfil}}%
     \else\def\@oddfoot{\hbox to \textwidth{\hfil\rm\thepage\hfil}}
     \fi
     \let\@evenfoot\@oddfoot
}

\def\@version#1{\ifnum\draftcontrol=1
\typeout{}\typeout{#1}\typeout{}
\vskip3mm\centerline{\hbox{\fbox{\normalsize{\tt DRAFT -- #1 -- }
                   {\draftdate}}}}\vskip3mm
\fi}
\let\version\@version
\long\def\eqlabel#1{\ifnum\draftcontrol=1
                    \tag@false  
                    \tag*{(\theequation) \hbox to -0.2cm{\hspace{0cm}\small{#1}\hss}}
                    \refstepcounter{equation}
                    \edef\@currentlabel{\theequation}
                    \ltx@label{#1}          
                    \else
                    \label{#1}
                    \fi
                    }
\let\st@bibitem\@bibitem
\let\st@lbibitem\@lbibitem
\ifnum\draftcontrol=1
  \def\@bibitem#1{%
    \st@bibitem{#1}\a@@label{#1}\ignorespaces}
  \def\@lbibitem[#1]#2{%
    \st@lbibitem[#1]{#2}\a@@label{#2}\ignorespaces}
  \def\a@@label#1{%
    \gdef\a@lab{\smash{\normalfont\small#1}}
    \ifvmode
      \if@inlabel
        \global\setbox\@labels\hbox{%
          \llap{\a@lab\let\a@lab\relax
                \kern\@totalleftmargin\kern\marginparsep}%
          \box\@labels}%
      \fi
    \fi}
\fi

\documentclass[12pt,letterpaper]{article}

\usepackage{amsmath,amssymb,array,calc,rotating,epsfig,psfrag}
\usepackage{tensind}  
\tensordelimiter{?}

\ifnum\draftcontrol=1
\fi

\renewcommand\baselinestretch{1.25}
\renewcommand\section{\@startsection {section}{1}{\z@}%
                                   {-3.5ex \@plus -1ex \@minus -.2ex}%
                                   {2.3ex \@plus.2ex}%
                                   {\normalfont\large\bfseries}}
\renewcommand\subsection{\@startsection{subsection}{2}{\z@}%
                                   {-3.25ex\@plus -1ex \@minus -.2ex}%
                                   {1.5ex \@plus .2ex}%
                                   {\normalfont\normalsize\bfseries}}
\renewcommand\subsubsection{\@startsection{subsubsection}{3}{\z@}%
                                   {-3.25ex\@plus -1ex \@minus -.2ex}%
                                   {1.5ex \@plus .2ex}%
                                   {\normalfont\normalsize\it}}
\renewcommand\paragraph{\@startsection{paragraph}{4}{\z@}%
                                   {-3.25ex\@plus -1ex \@minus -.2ex}%
                                   {1.5ex \@plus .2ex}%
                                   {\normalfont\normalsize\bf}}

\def\revise#1       {\raisebox{-0em}{\rule{3pt}{1em}}%
                     \marginpar{\raisebox{.5em}{\vrule width3pt\
                     \vrule width0pt height 0pt depth0.5em
                     \hbox to 0cm{\hspace{0cm}{%
                     \parbox[t]{4em}{\raggedright\footnotesize{#1}}}\hss}}}}

\def\calf         {{\cal F}}

\def\calh         {{\cal H}}

\def\call         {{\cal L}}

\def\sqr#1#2{{\vcenter{\vbox{\hrule height.#2pt
 \hbox{\vrule width.#2pt height#1pt \kern#1pt
 \vrule width.#2pt}\hrule height.#2pt}}}}

\def\a{\alpha}

\def\m{\mu}

\def\n{\nu}

\catcode`\@=12

\usepackage{color}

 \usepackage{amsmath}
 
\newcommand{\be}{\begin{equation}}
\newcommand{\ee}{\end{equation}}
\newcommand{\beq}{\begin{equation}}
\newcommand{\eeq}{\end{equation}}
\newcommand{\ba}{\begin{eqnarray}}
\newcommand{\ea}{\end{eqnarray}}

\def\lbldef#1#2{\expandafter\gdef\csname #1\endcsname {#2}}

\def\eqalign#1{\vcenter{\openup1\jot   }}

\def\href#1#2{#2}

\newcommand{\ber}{\begin{eqnarray}}
\newcommand{\eer}{\end{eqnarray}}

\newcommand{\bea}{\begin{eqnarray}}
\newcommand{\eea}{\end{eqnarray}}

\newcommand{\beqar}{\begin{eqnarray}}

\newcommand{\eeqar}{\end{eqnarray}}

\newcommand{\dsl}
  {\kern.06em\hbox{\raise.15ex\hbox{$/$}\kern-.56em\hbox{$\partial$}}}

\newcommand{\eeqarr}{\end{eqnarray}}
\newcommand{\ZZ}{{\rm \kern 0.275em Z \kern -0.92em Z}\;}

\def\a{\alpha}

\makeatletter \@addtoreset{equation}{section} \makeatother
\renewcommand{\theequation}{\thesection.\arabic{equation}}

\def\be{\begin{equation}}
\def\ee{\end{equation}}
\def\bea{\begin{eqnarray}}
\def\eea{\end{eqnarray}}
\def\m{\mu}
\def\n{\nu}

\def\ba{\bar{\alpha}}

\long\def\symbolfootnote[#1]#2{\begingroup%
\def\thefootnote{\fnsymbol{footnote}}\footnote[#1]{#2}\endgroup}

\usepackage{hyperref}

\begin{document}

\begin{titlepage}

\version\versionno

\vskip -.8cm

\rightline{\small{\tt MCTP-08-62}}

\vskip 1.7 cm

\centerline{\bf \Large L\"uscher Term for $k$-string Potential from }

\vskip 1 cm

\centerline{\bf \Large Holographic One Loop Corrections  }

\vskip .2cm \vskip 1cm {\large } \vskip 1cm

\centerline{\large Leopoldo A. Pando Zayas${}^1$\symbolfootnote[1]{lpandoz@umich.edu}, V. G. J. Rodgers${}^2$\symbolfootnote[2]{vincent-rodgers@uiowa.edu} and Kory Stiffler${}^2$\symbolfootnote[3]{corresponding author: kstiffle@gmail.com}}

\vskip .5cm
\centerline{\it ${}^1$Michigan Center for Theoretical
Physics}
\centerline{ \it Randall Laboratory of Physics, The University of
Michigan}
\centerline{\it Ann Arbor, MI 48109-1040}

\vskip .5cm
\centerline{\it ${}^2$Department of Physics and Astronomy}
\centerline{ \it  The University of Iowa}
\centerline{\it Iowa City, IA 52242}

\vspace{.8cm}

\begin{abstract}
{\small
We perform a systematic analysis of $k$-strings in the framework of
the gauge/gravity correspondence. We discuss the Klebanov-Strassler  supergravity background which is known to be
dual to a confining supersymmetric gauge theory with chiral symmetry breaking.
We obtain the $k$-string tension in
agreement with expectations of field theory.
Our main new result is the study of one-loop corrections on the string theoretic side.  We explicitly find the frequency spectrum for both the bosons and the fermions for quadratic fluctuations about the classical supergravity solution.  Further we use the massless modes to compute $1/L$ contributions to the one loop corrections to the $k$-string energy. This corresponds to the L\"uscher term contribution to the k-string potential on the gauge theoretic side of the correspondence.}

\end{abstract}



\end{titlepage}



\section{Introduction }
Although the QCD Lagrangian has been known for more than forty years, extracting physical predictions from it in the strong coupling regime has proven to be monumental.  Ingenious string theoretic strategies have been devised since that time that attempt to probe the theory analytically and provide insight into QCD.  The Wilson loop,
\[ <W(\mathcal{C})> = \exp{(-T \mathcal{A}_\mathcal{C})} \] for example, where $T$ is the string tension and $\cal{A}$ is the area circumscribed by the contour ${\cal C}$, computes the probability amplitude that quarks will traverse the contour and has a string theoretic interpretation in terms of gluon flux tubes.  In 2+1 dimensions, for example, Karabali, Kim and Nair (KKN) \cite{Karabali:1998yq} were able to compute the string tension starting from first principles in SU(N) Yang-Mills and found that
\[T_F = e^4\frac{(N^2-1)}{8 \pi},  \] for fermions in the fundamental representation. This result is within $3\%$ agreement of earlier theoretical predictions from lattice gauge theory work by first Teper \cite{Teper:1998te} and then Lucini, Teper and Wegner \cite{teper}. More recent work, however has suggested that in the $N\rightarrow \infty$ limit, the lattice data is about $1\%$ lower than the work of KKN \cite{Bringoltz:2006zg}.  This residual $1\%$ remains even after all systematics are taken into account.  However Karabali, Yelnikov, and Nair are presently improving the wavefunction in KKN and preliminary results for the first order corrections have moved the agreement to within $1-2\%$ for fixed values of $N$ and $.88\%$ for $N\rightarrow \infty$ \cite{nair}.  

The AdS/CFT correspondence \cite{mn} goes further and opens a
window into the strong coupling region of field theory by means of supergravity backgrounds. As shown in Table[\ref{tab:GaugeTHeoryStatesAndTheirStringTheoryConfigurations}] we now have the correct string theory correspondence for specific gauge field theory states\cite{pandoz}.
\begin{table*}[hbp]
	\centering
			\begin{tabular}{|c|c|}\hline
\multicolumn{2}{|c|}{ \bf Summary of AdS/CFT Correspondences} \\ \hline\hline
\textbf{Gauge Theory State} & \textbf{String Theory Configuration}  \\ \hline \hline
Glueballs & Spinning Folded Closed String  \\ \hline
Mesons of heavy quarks & Spinning open strings ending on boundary\\ \hline
Baryons of heavy quarks & Strings attached to baryonic vertex \\ \hline
Dibaryons & Strings attached to wrapped branes \\ \hline
Mesons of light quarks & Spinning open strings ending on D7 branes \\ \hline
	$k$-strings & Wrapped branes with flux\\ \hline	
		\end{tabular}
	\caption{Gauge Theory States and Their String Theory Configurations}
	\label{tab:GaugeTHeoryStatesAndTheirStringTheoryConfigurations}
\end{table*}
Our focus will be on the k-string configurations in confining field theories as they have received a lot of attention as prototypes of the AdS/CFT probe. The $k$-string is the flux tube that results between $k$ quarks in the fundamental and $k$ antiquarks. These configurations have been studied following various methods. Earlier analytical results were obtained in Douglas-Shenker \cite{ds} and the MQCD result of
Hanany, Strassler and Zaffaroni  \cite{hsz}. $k$-strings are good candidates for examing Ads/CFT since they are also an active line of research in
lattice gauge theories \cite{teper, deldebbio, teper1, teper2}. One of the issues that these configurations are able to probe is whether there is a Casimir-like  scaling \[ T_k\approx k(1-k/N) \] for large N or whether the scaling exhibits a sine law where \[ T_k \propto N \sin{(\frac{\pi k}{N})}.\]  What adds to the controversy is that on the one hand, the $1/N$ expansion of QCD agrees with the sine law scaling  \cite{Armoni:2003ji,Armoni:2003nz}, while on the other hand lattice calculations favor Casimir scaling ($1/N$) \cite{Bringoltz:2008nd}.  Nevertheless, both the sine law and the Casimir law respect the symmetry $k \to N - k$, an important feature of $k$-strings which describes replacing quarks with anti-quarks~\cite{hk}. 

There is, by now, a vast literature of $k$-strings in the context of the gauge/gravity correspondence that originates with this and other interesting observations made in  \cite{hk,h} and extends to investigations of the $k$-string width due to the electric flux \cite{Polchinski:2001ju,Armoni:2008sy}.  A very complete review is presented by Shifman in \cite{shifman}.
Our goal is to revisit the $k$-string tension in the context of the gauge/gravity correspondence and  more importantly to understand the structure of fluctuations and their contributions to the L\"uscher term. The $k$-string tension can be computed as the classical value of the Hamiltonian for a particular static classical supergravity  configuration.  In this work we start with the Born-Infeld action associated with a  Klebanov-Strassler \cite{ks} supergravity background and evaluate the tension from the solutions of the field equations making no approximations.  We find, in agreement with \cite{hk}, that tension is \[\frac{T_k}{T_f} \approx \sin{(\frac{\pi k}{N})}. \] We also find that it \emph{exactly} satisfies the aforementioned $k \to N  - k $ symmetry.
Our main result of this note is to compute the quadratic fluctuations of the Hamiltonian incorporating both bosonic and fermion fields. These fluctuations correspond to the one-loop correction to the $k$-string energy.  By summing the zero point energies we are able to calculate the leading order corrections to the energy as well as the  L\"uscher \cite{Luscher:1980fr} term, \[V_{\mbox{\small{L\"uscher}}} = -\frac{\pi}{3L}. \]  Note, that since the L\"uscher term in independent of $k$, it too satisfies the important $k \to N - k$ symmetry.

\section{Electrically and magnetically charged Dp branes}\label{general}
In this section we discuss the general formalism. This is a classical analysis that can be found
in various papers, we present it here for completeness. The starting point is the action of a Dp brane in
a supergravity background

\be
\begin{split}
&S_{Dp}=-\m_p\int d^{p+1} \xi~e^{-\Phi}\, \sqrt{-{\rm det}\left(g_{\m\n}+B_{\m\n}+2\pi \a'F_{\m\n}\right)}\\
&\qquad \qquad +\m_p\int \exp(\mathcal{F}_2)\wedge\sum\limits_q C_q. \label{dp}
\end{split}
\ee
In this expression $g_{\m\n}$ and $B_{\m\n}$ are the induced metric and B-field in the world volume of the
Dp brane. The field strength $F_{\m\n}$ describes a $U(1)$ potential which in turns induces the
electric and magnetic charges in the world volume of the Dp brane. Our problem can basically be formulated as:
{\it given a supergravity background find classical solutions of the embedding describing electrically and/or magnetically
charged Dp branes; compute the energy of such configurations and the spectrum of its small fluctuations. }

From the gauge theory point of view we describe a string extending along the coordinates $(t,x)$.  From the gravity side we are going to describe a D3 brane wrapping a two-cycle which we parameterize as $(\theta, \phi)$.  Therefore the world volume coordinates are
\be
  \xi^{\m} = (t, x, \theta, \phi).
\ee
We also turn on a gauge field in the world volume of the brane which is described by two non-vanishing components of the field strength, $F_{tx}$ and $F_{\theta\phi}.$  Thus
\be
g_{\m\n}+B_{\m\n}+2\pi \a'F_{\m\n} =
\ee
\be
\left(
\begin{array}{cccc}
-g_{tt} &2\pi\alpha' F_{tx}&0&0\\
-2\pi\alpha' F_{tx} &g_{xx}&0&0\\
0&0&g_{\theta\theta}& g_{\theta\phi}+ B_{\theta\phi}+2\pi\alpha' F_{\theta\phi}\\
0&0&g_{\theta\phi}-B_{\theta\phi}-2\pi\alpha' F_{\theta\phi}& g_{\phi\phi}\\
\end{array}
\right).
\ee
The action for D3 branes is
\be \label{eq:D3action}
S=-\mu_3\int d^4\xi \sqrt{ -\det (g_{\m\n} + {\cal F}_{\m\n})} + 2\pi \alpha' \mu_3 \int  {\cal F}_2 \wedge C_2.
\ee

\bea
S&=&-\mu_3 \int d^4 \xi e^{-\Phi}\sqrt{g_{tt}g_{xx}-(2\pi\alpha' F_{tx})^2}\sqrt{g_{\theta\theta}g_{\phi\phi}-g_{\theta\phi}^2 + \calf_{\theta\phi}^2}
\nonumber \\
&+& 2\pi \alpha' \mu_3 \int d^4 \xi F_{tx}(C_2)_{\theta\phi}.
\eea

One can consider the equation of motion for $F_{tx}$ and find that
\be
\frac{\partial \call }{\partial F_{tx}}=D={\rm const.}
\ee
where $D$ is the displacement.
A way to determine the constant is through the quantization conditions which arises due to the
coupling of the B-field \cite{reyfluc,alfonso,alfonsofluctuations}:
\be
\int\limits_{S^2} d^2\xi \frac{\partial \call }{\partial F_{tx}}=\frac{p}{2\pi \alpha'}.
\ee

The situation for the magnetic component is more straightforward, one simply demands
$F_{\theta\phi}$ to be quantized \cite{bachas}
\be
F_{\theta\phi}=-\frac{q}{2} \sin\theta d\theta \wedge d\phi.
\ee
This leads to $q$ units of flux after integration over the 2-sphere which the D3 brane wraps.

One can now define the Hamiltonian density as
\be
\calh =DF_{tx}-\call.
\ee

The Hamiltonian takes the form
\be
H=\mu_3\int\limits_{\mathbb{R}\times S^2}d^3\xi \sqrt{g_{tt}g_{xx}}
\sqrt{e^{-2\Phi}(g_{\theta\theta}g_{\phi\phi}-g_{\theta\phi}^2+\calf_{\theta\phi}^2)+(D-C_2)^2}.  \label{Hamiltonian}
\ee

\section{K-Strings in the Klebanov-Strassler background}\label{KS}

\subsection{Review of the Klebanov-Strassler background}
There are many interesting reviews of the Klebanov-Strassler background \cite{hk,Herzog:2002ih,Herzog:2001xk}. However to facilitate the reader, this section will review the salient features which are relevant to this work.
We begin by considering a collection of  $N$  regular
and $M$ fractional D3-branes in the geometry  of the deformed conifold
\cite{Minasian:1999tt, Ohta:1999we,Herzog:2001xk}.  The 10-d metric is of the form:

\be  ds^2_{10} =   h^{-1/2}(\tau)   ds_4^2  +  h^{1/2}(\tau) ds_6^2
\ ,  \ee  where $ds_4^2$ is the 4-D Minkowski metric and $ds_6^2$ is the metric of the deformed conifold
\cite{candelas,ks}:

\be
\label{mtmetric}
ds_6^2 = \frac{1}{2}\varepsilon^{4/3} K(\tau)  \Bigg[ \frac{1}{3
K^3(\tau)} (d\tau^2 + (g^5)^2)  +  \cosh^2 \left(\frac{\tau}{
2}\right) [(g^3)^2 + (g^4)^2]  + \sinh^2 \left(\frac{\tau}{
2}\right)  [(g^1)^2 + (g^2)^2] \Bigg].
\ee
where
\be
K(\tau)= \frac{ (\sinh (2\tau) -
2\tau)^{1/3}}{ 2^{1/3} \sinh \tau},
\ee
and
\bea
\label{forms}
g^1 &=&
\frac{1}{ \sqrt{2}}\big[- \sin\theta_1 d\phi_1  -\cos\psi\sin\theta_2
d\phi_2 + \sin\psi d\theta_2\big] ,\nonumber \\  g^2 &=& \frac{1}{
\sqrt{2}}\big[ d\theta_1-  \sin\psi\sin\theta_2 d\phi_2-\cos\psi
d\theta_2\big] , \nonumber \\  g^3 &=& \frac{1}{\sqrt{2}} \big[-
\sin\theta_1 d\phi_1+  \cos\psi\sin\theta_2 d\phi_2-\sin\psi d\theta_2
\big],\nonumber \\  g^4 &=& \frac{1}{ \sqrt{2}} \big[ d\theta_1\ +
\sin\psi\sin\theta_2 d\phi_2+\cos\psi d\theta_2\ \big],   \nonumber
\\  g^5 &=& d\psi + \cos\theta_1 d\phi_1+ \cos\theta_2 d\phi_2.
\eea

The 3-form fields are:
\begin{eqnarray}
F_3 = dC_2 &=& \frac{M\alpha'}{ 2} \left \{g^5\wedge g^3\wedge g^4 + d [
F(\tau)  (g^1\wedge g^3 + g^2\wedge g^4)]\right \} \nonumber \\  &=&
\frac{M\alpha'}{ 2} \left \{g^5\wedge g^3\wedge g^4 (1- F)  + g^5\wedge
g^1\wedge g^2 F \right. \nonumber \\  && \qquad \qquad \left. + F'
d\tau\wedge  (g^1\wedge g^3 + g^2\wedge g^4) \right \}\ ,
\end{eqnarray}

\noindent and
\begin{eqnarray}
H_3 = dB_2 &=& \frac{g_s M \alpha'}{ 2} \bigg[  d\tau\wedge (f'
g^1\wedge g^2  +  k' g^3\wedge g^4)  \nonumber \\  && \left. + \frac{1}{
2} (k-f)  g^5\wedge (g^1\wedge g^3 + g^2\wedge g^4) \right]\ .
\end{eqnarray}

\noindent Solving these for $B_2$ and $C_2$ results in
\begin{align}\label{Bfield}
B_2 &= \frac{g_s M \alpha'}{ 2} [f(\tau) g^1\wedge g^2  +
k(\tau) g^3\wedge g^4 ] \\
\label{C2field}
C_2 &= \frac{M}{8 \pi T_0}[ 2 F(\tau)(g^1 \wedge g^3 + g^2\wedge g^4) + (\cos\psi \sin\theta_1 \sin\theta_2 -\cos\theta_1 \cos\theta_2)d\phi_1\wedge d\phi_2 \nonumber\\
&~~~~- \cos\psi d\theta_1\wedge d\theta_2
        +\psi(\sin\theta_1 d\theta_1\wedge d\phi_1- \sin\theta_2 d\theta_2\wedge d\phi_2)\nonumber\\
        &~~~~-\sin\psi \sin\theta_1 d\phi_1\wedge d\theta_2 + \sin\psi \sin\theta_2 d\phi_2\wedge d\theta_1]
\end{align}

The self-dual 5-form field strength is  decomposed as $\tilde F_5 =
{\cal F}_5 + \star {\cal F}_5$, with  \be  {\cal F}_5 = B_2\wedge F_3
= \frac{g_s M^2 (\alpha')^2}{4} \,\ell(\tau)  g^1\wedge g^2\wedge
g^3\wedge g^4\wedge g^5\ ,  \ee  where  \be  \ell = f(1-F) + k F\ ,
\ee  and  \be  \star {\cal F}_5 = 4 g_s M^2 (\alpha')^2
\varepsilon^{-8/3}  dx^0\wedge dx^1\wedge dx^2\wedge dx^3  \wedge
d\tau \frac{\ell(\tau)}{ K^2 h^2 \sinh^2 (\tau)}\ .  \ee  The functions
introduced in defining the form fields are:  \bea  F(\tau) &=& \frac{\sinh
\tau -\tau}{ 2\sinh\tau}\ ,  \nonumber \\  f(\tau) &=&
\frac{\tau\coth\tau - 1}{ 2\sinh\tau}(\cosh\tau-1) \ ,  \nonumber \\
k(\tau) &=& \frac{\tau\coth\tau - 1}{ 2\sinh\tau}(\cosh\tau+1)  \ .
\eea  The equation for the warp factor is  \be \label{firstgrav}  h' =
- \alpha \frac{f(1-F) + kF}{ K^2 (\tau) \sinh^2 \tau}  \ ,  \ee  where
\be
\alpha =4 (g_s M \alpha')^2  \varepsilon^{-8/3}\ .
\ee
For large $\tau$ we impose the boundary condition that  $h$
vanishes. The resulting integral expression for $h$ is
\be
\label{intsol}  h(\tau) = \alpha \frac{ 2^{2/3}}{ 4} I(\tau) =  (g_s
M\alpha')^2 2^{2/3} \varepsilon^{-8/3} I(\tau)\ ,
\ee
where
\be
I(\tau) \equiv  \int_\tau^\infty d x \frac{x\coth x-1}{ \sinh^2 x}
(\sinh (2x) - 2x)^{1/3}  \ .  \ee
The above integral has the
following expansion in the IR:
\be  I(\tau\to 0) \to I_0 - I_1
\tau^2 + {\cal O}(\tau^4) \ ,
\ee
where $I_0\approx 0.71805$ and
$I_1=2^{2/3}\, 3^{2/3}/18$. The absence  of a linear term in $\tau$
reassures us that we are really expanding  around the end of space,
where the Wilson loop will find it more favorable to arrange itself.

\subsection{Review of Herzog-Klebanov $k$-String Tension}

In this section we review the original calculation presented
by Herzog and Klebanov (HK) in \cite{hk}. Our
goal is to clarify the extent of their simplifications and provide a more complete computation of the strings tension.

Herzog and Klebanov construct their model by considering the metric in the $\tau=0$ limit.  By performing an S-duality transformation they are able to send $F_3 \leftrightarrow H_3$ and exchange $k$ fundamental strings for $k$ D1 branes.  Further a T-duality transformation along the D1-brane direction yields $k$ D0-branes on an $S^3$ with $M$ units of NS-NS flux.

This is related to the setup of Bachas, Douglas and Schweigert \cite{bachas} who
found that $q$ D0 branes blow up into an $S^2$ via the Myers effect.
We consider a D3-brane wrapping the $S^2$, that is with world volume given by the parametrization
\begin{align}\label{eq:parametrization}
  X^0 = t,~X^1 = x,~\theta_1 = \theta,~\phi_2 = -\phi.
\end{align}

\noindent The KS background metric at
\begin{align}\label{eq:embedding}
  X^2 = X^3 = \tau = 0,~\theta_2 = \theta,~\phi_1 = \phi
\end{align}
\noindent becomes
\begin{align}
ds_{10}^2 &= h_0^{-1/2}ds_4^2
+ b M \a'(\frac{1}{4}d\psi^2 + \cos^2\frac{\psi}{2}(d\theta^2 + \sin^2 \theta d\phi^2)) \nonumber\\
   b &= 2^{2/3}3^{-1/3}I_0^{1/2} \approx 0.933
\end{align}

\noindent where the new constants $h_0$ and $K_0$ are the $\tau \to 0$ limits of $h(\tau)$ and $K(\tau)$, respectively.

After S-duality we find that, for $\tau=0$ we have  $F_5=0$, and $C_2=0$. Under S-duality the RR 3-flux becomes
$H_3$ flux.  The corresponding $B$-field is given by
\be
B_2 = \frac{\alpha' M}{2} \left(\psi + \sin\psi \right) \sin\theta d\theta \wedge d\phi.
\ee

There is an ambiguity in choosing this $B$ field. The quantity that appears in the equation of motion is
\bea
H_3&=&dB_2=  \frac{\alpha' M}{2} \left(1 + \cos \psi \right) d\psi \wedge d\theta \wedge \sin \theta d\phi \nonumber \\
\eea

By applying a coordinate change $\psi \to 2\psi - \pi$, the metric KS background at $\tau=0$ becomes
\be
h_0^{-1/2}ds_4^2
+ b M \a'(d\psi^2 + \sin^2\psi(d\theta^2 + \sin^2 \theta d\phi^2)).
\ee

\noindent and $H_3$ becomes
\bea
H_3&=&dB_2=  \alpha' M \left(1 - \cos 2\psi \right) d\psi \wedge d\theta \wedge \sin \theta d\phi \nonumber \\
&=&2 \alpha' M \sin^2\psi d\psi \wedge d\theta \wedge \sin \theta d\phi.
\eea

\noindent leading to a $B_2$ of the form
\be
B_2 = \alpha' M \left(\psi -\frac12 \sin2\psi \right) \sin\theta d\theta \wedge d\phi.
\ee

We will also consider a world volume $U(1)$ field strength
\be
F=-\frac{q}{2}\sin\theta d\theta\wedge d\phi.
\ee
This is a magnetically charged $U(1)$ field. Naturally, it represents the charges of D1-strings.  The charge $q$ represents electrically charged strings under an S-duality transformation which are further
interpreted as $q$ quarks on the gauge theory side.

Since for this case $C_4=C_2=0$ we have no contribution from the Chern-Simon term.
The action is then
\be
S=T_3\int dt dx d\theta d\phi \, h_0^{-1/2}\, \sin\theta \, M \a'
\sqrt{b^2 \sin^4\psi + \left(\psi -\frac{\sin2\psi}{2}-\frac{\pi\, q}{M}\right)^2},
\ee

Minimizing the Hamiltonian with respect to $\psi$ one finds:

\be
\psi-\frac{\pi\, q}{M}=\frac{1-b^2}{2}\sin2\psi.
\ee
By substituting the solution into the Hamiltonian one finds the $k$-string tension
\begin{align}
T&\approx b\sin\psi \sqrt{1+(b^2-1)\cos^2\psi}.
\end{align}
Herzog and Klebanov showed that since $b\approx 1$ one obtains that
\begin{align}
T_q &\sim b\sin\frac{\pi \, q}{M}.
\end{align}

\subsection{The k-string tension in the KS background}\label{tension}
We propose that under the parametrization~\ref{eq:parametrization}, the D-Brane action has a solution given by $\psi= constant$ and equation~\ref{eq:embedding}.  We will demonstrate this by directly solving the field equations for the bosonic fields, and later will show that it is indeed a classical solution when we fluctuate around it.  Then no first order fluctuations survive up to total derivative terms.  Following the strategies of ~\cite{Ridgway:2007vh,tye}, we proceed to evaluate the tension without applying S-duality. 

In the KS background, we find the pullbacks of $B_2$ and the Ramond-Ramond fields $C_i$ to be
\begin{align}
  B_2 &= C_0 = C_4 = 0 \nonumber\\
  C_2 &=\frac{M\alpha'}{2}(\psi + \sin\psi)\sin\theta d\theta\wedge d\phi.
\end{align}
Further we consider electrically charged D3-branes, along the $x$-direction so that
\begin{align}
  F_{tx} = \mbox{constant,}
\end{align}
and all other components are zero. Normally due to the fibration of 10D space-time one does not get a round metric as the induced metric.  Our result relies on a very specific choice  of the volume coordinates.  We find the induced metric to have topology $R^{1,1} \times S^2$:
\begin{align}\label{eq:ginducedsimplified}
  ds^2 &=  g_{\mu\nu}d\xi^{\mu}d\xi^{\nu} = h_0^{-1/2}(-dt^2 + dx^2) + \frac{2}{R}(d\theta^2 + \sin^2\theta d\phi^2)
\end{align}

\noindent where the scalar curvature is
\begin{equation}
  R  = \frac{2 \sec^2 \frac{\psi}{2}}{h_0^{1/2} \varepsilon^{4/3} K_0}.
\end{equation}

From the action,~Eq.(\ref{eq:D3action}), we explicitly calculate the field equations for $\tau$, $\psi$ and $A_x$ and find that when $\tau \rightarrow 0$ and $\psi=\psi[\theta]$ the field equation for $\psi$ reduces to
\begin{align}
  (b_1 f_3(\psi,\psi',\psi'') - f_1(\psi,\psi') f_2(\psi,\psi')) \sin(\theta) + b_1 f_2(\psi,\psi') \cos(\theta) \psi'(\theta) = 0
\end{align}

\noindent where
\begin{align}
  f_1(\psi,\psi')^2 &= {a_1}^2 b_1 \left(1 + \cos(\psi(\theta))\right)\left(2 + 2 \cos(\psi(\theta)) + \psi'(\theta)^2\right) \\
  f_2(\psi,\psi') &= 4 \cos^2\left(\frac{\psi(\theta)}{2}\right) + \psi'(\theta)^2 \\
  f_3(\psi,\psi',\psi'') &= 8 \cos^2\left(\frac{\psi(\theta)}{2}\right) \sin(\psi(\theta)) + 3 \sin(\psi(\theta)) \psi'(\theta)^2 + \nonumber\\
                   &~~~~~~~~~~~~~~~~~~~~~~~~~~~~~~~~~~~+ 4 \cos\left(\frac{\psi(\theta)}{2}\right)^2 \psi''(\theta) \\
  a_1 &= 2^{1/6}3^{1/3} F_{tx} g_s M,~~~b_1 = 4\pi^2 T_0^2 \epsilon^{8/3}(T_0^2 - F_{tx}^2 h_0),~~~T_0 = \frac{1}{2\pi\a'}.
\end{align}

By exploring the solution $\psi[\theta]=\psi_0,$ a constant, the field equation for $\psi$ reduces to
 \begin{equation}
    b^2 T_0^2 \tan^2\left(\frac{\psi_0}{2}\right) = F_{tx}^2 h_0 \left(1 + b^2 \tan^2\left(\frac{\psi_0}{2}\right) \right)
\end{equation}

The $\psi_0$ dependence of $F_{tx}$ can be further obtained from the Euler-Lagrange equations for $A_x$.  Since
\[ \frac{\partial \mathcal{L}}{\partial A_{x}} =0,\]
define $D$ as
\begin{align}
 D \equiv \frac{\partial \mathcal{L}}{\partial F_{tx}} = \mbox{constant}.
\end{align}
One can solve for the dependence of $\psi_0$ and write
\begin{equation}
F_{tx}= \frac{T_0^2 (D-\Omega )}{\sqrt{h_0} \sqrt{\Delta^2 + (T_0(D - \Omega))^2}},
   \end{equation}
where
\begin{eqnarray}
&\Delta = \frac{T_0^2}{2\pi g_s} \int d\theta d\phi\,\frac{2}{R}\sin\theta = \frac{4 T_0^2}{R g_s}\nonumber\\
 &\Omega = \frac{T_0}{2\pi} \int d\theta d\phi C_{\theta\phi} = \frac{M}{2 \pi}(\psi_0 + \sin\psi_0\textit{}).
 \end{eqnarray}
The solution to the field equations requires that  $\psi_0$ satisfies
\bea\label{eq:psicondition}
   \psi_0 - \frac{2D\pi}{M} &=& (b^2 - 1)\sin\psi_0
\eea where
$b=\frac{2^{2/3} \sqrt{I_0}}{\sqrt[3]{3}}.$ This constraint equation from minimizing the action is the same that one arrives at when minimizing the Hamiltonian density:
\begin{align}
  \mathcal{H} &= D F_{tx} - \mathcal{L} =  h_0^{-1/2}\sqrt{\Delta^2 + T_0^2(D - \Omega)^2}
\end{align}
with respect to $\psi_0.$
Evaluating $\mathcal{H}$ at $\psi_0$ yields the minimized Hamiltonian:
\begin{align}
\mathcal{H}_0 &= \frac{bMT_0}{h_0^{1/2} \pi} \cos\frac{\psi_0}{2} \sqrt{1 + (b^2 - 1) \sin^2\frac{\psi_0}{2}}.
\end{align}
We interpret this as the $k$-string tension. Since $b\sim 1$, equation~\ref{eq:psicondition} yields that $\psi_0 \approx \frac{2D\pi}{M}$.  Upon setting $D = k - M/2$, the $k$-string tension \emph{approximately} simplifies to the \emph{sine} law:
\begin{equation}
 T_k \approx \frac{bMT_0}{h_0^{1/2} \pi} \sin \left(\frac{k\pi}{M}\right)
\end{equation} which vanishes when $k$ is an integer multiple of $M$.

\begin{figure}
  \centering
  \includegraphics[width=0.45\columnwidth]{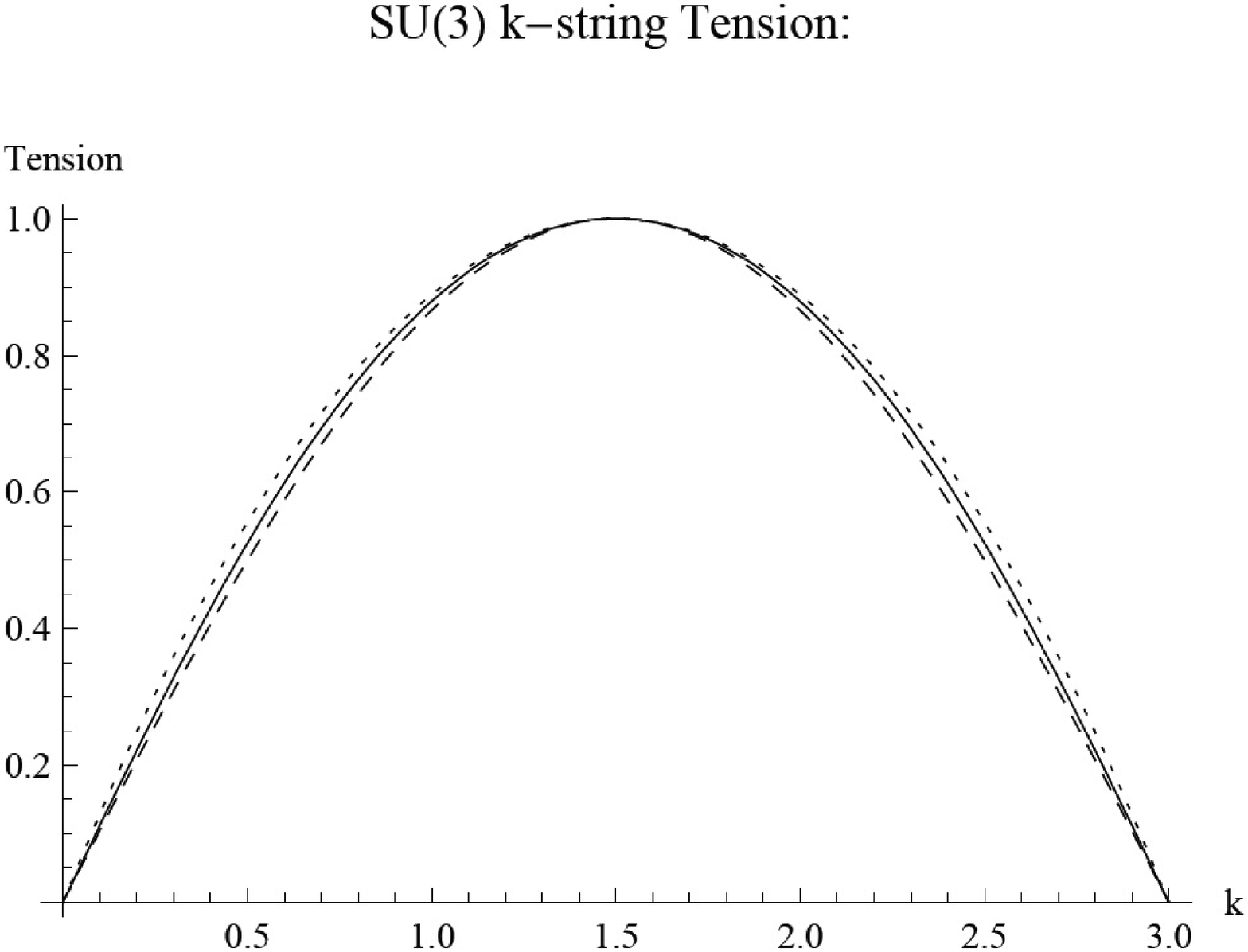}
  \hfill
  \includegraphics[width=0.45\columnwidth]{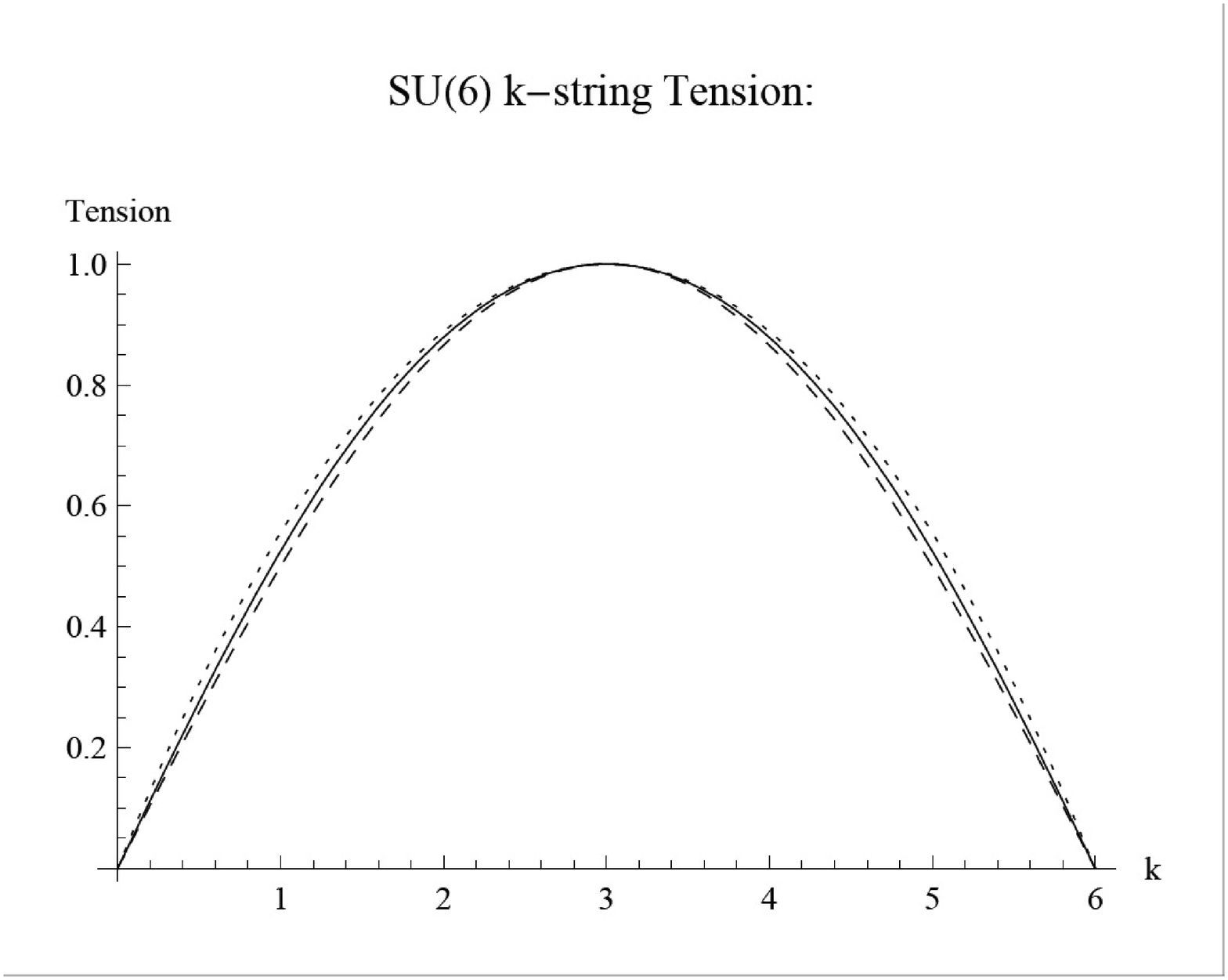}
  \caption{Exact solutions for the $k$-string tension for M=3 and M=6, normalized to one, and compared to the sine law scaling and Casimir scaling. Solid: Exact numerical solution.   Dashed: sine law.  Dotted: Casimir Law.  We find similar agreement for larger M.}
  \label{fig:kstringtension}
\end{figure}

We can numerically solve the transcendental equation for $\psi_0$ for given $M$, and plot the resulting $k$-string tension as a function of $k$.  Figure~\ref{fig:kstringtension} compares the exact $k$-string tension to sine law scaling and Casimir scaling for $M=3$ and $M=6$.  The exact KS tension sits between the sin law and the Casimir law. 

So in our analysis, we find approximate agreement with the sine law from lattice gauge theories of QCD.  The SU(M) $k$-string tension vanishes when the D3-branes flux, $D=k-M/2$, is quantized by an integer or half integer multiple of a M.  Thus in our specific example, the gauge/gravity correspondence is manifested specifically between SU(M) $k$-strings and D3-branes endowed with electric flux in the Klebanov-Strassler background.

\section{Excitations of $k$-strings}
Up until now, we have only calculated the ground state \emph{k}-string energy, $E_0 = T_k L$.  We now calculate one loop quantum corrections to this energy by allowing the fields to fluctuate around their classical solutions, expecting a Luscher term, proportional to $1/L$.  The vanishing of first order contributions in the fluctuating fields assures us that we are fluctuating around a classical field configuration.

\subsection{Fluctuations Around the Classical Solutions of the D3-brane action}
We proceed by investigating small fluctuations around the classical solutions of the bosonic fields $X^a$, the gauge potentials, $A_{\mu}$, and the fermion fields, $\Theta$~\cite{bachas}, of the D3-brane action:
\begin{align}\label{eq:shift}
            X^a =& X^a_{(0)} + \delta X^a(\xi), \nonumber\\
            A^{\mu} =& A^{\mu}_{(0)} + \delta A^{\mu} (\xi), \nonumber\\
            \Theta =& 0 + \delta\Theta(\xi).
\end{align}
The equations of motion for these new fields will yield the energy eigenvalues of the Hamiltonian for the quadratic fluctuations.

The general form of the action for D3-branes breaks up into bosonic and fermionic terms~\cite{bt,m1,m2,m3,m4,ingo}
\begin{align}
  S &= S^{(b)} + S^{(f)},
\end{align}
where the bosonic action $S^{(b)}$ is the same as $S_{(Dp)}$ in Eq.[\ref{dp}] while
$S^{(f)}$, is
\begin{align}\label{eq:Sfexpanded}
    S^{(f)} &= \frac{T_0^2}{4\pi g_s}\int d^4\xi e^{\Phi}\sqrt{-|M_0|}\overline{\Theta}[\left(M_0^{-1}\right)^{\alpha\beta}\Gamma _{\alpha}\partial_{\beta}+M_1+M_2+M_3]\Theta.
\end{align}

\noindent Here, $M_0$, is a sum of the classical values of $g^{(0)}$ and $\mathcal{F}^{(0)}$
\begin{align}\label{eq:M0}
M_0 &= g^{(0)} + \mathcal{F}^{(0)},
\end{align}
and $M_1$, $M_2$, $M_3$, and the $\Gamma^{\alpha}$ matrices are calculated in appendix~\ref{app:fermions}.

We proceed by using  the reparametrization invariance of the D3-brane world volume  to reduce our considerations to only six fluctuations out of the 10 bosonic SUGRA fields. The world volume coordinates take the static gauge of
\begin{equation}
 X^0 = t,~~~X^1 = x,~~~\theta_p \equiv \frac{1}{2}(\theta_1 + \theta_2) = \theta,~~~\phi_m \equiv \frac{1}{2}(\phi_1 - \phi_2) = \phi.
 \end{equation}
We vary the fields from their classical solutions in the following way,
\begin{align}\label{eq:bosonicfluctuations}
   \theta_m &\equiv \frac{1}{2}(\theta_1 - \theta_2) = \delta\theta_m,~~~\phi_p \equiv \frac{1}{2}(\phi_1 + \phi_2) = \delta\phi_p \nonumber\\
   X^2 &= \delta X^2,~~~X^3 = \delta X^3 \nonumber\\
  \psi &= \psi_0 + \delta\psi,~~~\tau = \tau_0 + \delta\tau,
\end{align}

\noindent where we will be careful to take the $\tau_0 \to 0$ limit at the appropriate time to avoid singularities.

With this explicit shift, the fields $B_{ab}, C_{ab}$, and the 10-D bosonic metric, $G_{ab}$, all can be expanded to at least quadratic order in the fluctuations:
\begin{align}\label{eq:BulkExpansion}
  G_{ab} &= G_{ab}^{(0)} + G_{ab}^{(1)} + G_{ab}^{(2)} \nonumber\\
  B_{ab} &= B_{ab}^{(1)} + B_{ab}^{(2)} \nonumber\\
  C_{ab} &= C_{ab}^{(0)} + C_{ab}^{(1)} + C_{ab}^{(2)}.
\end{align}

\noindent The Ramond-Ramond four-form has only one non-vanishing component, which is second order in the fluctuations in the $\tau_0 \to 0$ limit
\begin{equation}
  C_4 \propto \lambda^2\delta\tau^2~dX^0 \wedge dX^1 \wedge dX^2 \wedge dX^3,
\end{equation}

\noindent and so is zero to quadratic order when pulled back to the D3-brane.

The $U(1)$ gauge potential undergoes a very simple shift, and is only first order in the fluctuations:
\begin{align}
  F_{\mu\nu} &= F_{\mu\nu}^{(0)} + \delta F_{\mu\nu} \nonumber\\
             &= (\partial_{\mu}A_{\nu}^{(0)} - \partial_{\nu}A_{\mu}^{(0)}) + (\partial_{\mu}\delta A_{\nu} - \partial_{\nu} \delta A_{\mu}) \nonumber\\
             &= F_{tx}^{(0)} (?\delta^t_\mu??\delta^x_{\nu}? - ?\delta^t_\nu??\delta^x_\mu?) + (\partial_{\mu}\delta A_{\nu} - \partial_{\nu} \delta A_{\mu})
\end{align}

The Chern-Simons part of the D-brane action is, to quadratic order,
\begin{align}\label{eq:chernsimonsexpanded}
   S_{cs} &= \frac{T_0^2}{2\pi} \int \exp(\mathcal{F})\wedge\sum\limits_q C_q = \frac{T_0^2}{2\pi} \int \mathcal{F}\wedge C_2 \nonumber\\
          &= \frac{T_0^2}{2\pi} \int [\mathcal{F}^{(0)} \wedge C_2^{(0)} + (\mathcal{F}^{(1)} \wedge C_2^{(0)} + \mathcal{F}^{(0)} \wedge C_2^{(1)}) + \nonumber\\
          &~~~~+(\mathcal{F}^{(0)} \wedge C_2^{(2)} + \mathcal{F}^{(1)} \wedge C_2^{(1)} + \mathcal{F}^{(2)} \wedge C_2^{(0)})].
\end{align}

We now expand the square root in the bosonic action to quadratic order.  We will utilize the following formula to accomplish this:
\begin{align}
\sqrt{|M_0 + \delta M_0|} = \sqrt{|M_0|}&\left\{1 + \frac{1}{2} \mbox{Tr}(M_0^{-1} \delta M_0) + \frac{1}{8} [\mbox{Tr}(M_0^{-1} \delta M_0)]^2 + \right.\nonumber\\
                             &\left.-\frac{1}{4} \mbox{Tr}(M_0^{-1} \delta M_0 M_0^{-1}\delta M_0) + \mathcal{O}(\delta M_0^3)\right\},
\end{align}

\noindent where
\begin{align}
  \delta M_0 &= M_0^{(1)} + M_0^{(2)} \nonumber\\
   M_0^{(1)} &= g^{(1)} + \mathcal{F}^{(1)},~~~M_0^{(2)} = g^{(2)} + \mathcal{F}^{(2)}.
\end{align}

\noindent Putting all this together, we find the bosonic action to be, to quadratic order,
\begin{align}\label{eq:bosonicactionexpanded}
  S^{(b)} &= S_{(0)}^{(b)} + S_{(1)}^{(b)} + S_{(2)}^{(b)},
\end{align}

\noindent where the first and second order actions have the following forms:
\begin{align}\label{eq:1storderbosons}
  S_{(1)}^{(b)} &= -\frac{T_0^2}{2\pi g_s} \int d^{4}\xi \sqrt{-|M_0|} \frac{1}{2}\mbox{Tr}(M_0^{-1} M_0^{(1)}) + \frac{T_0^2}{2\pi} \int [\mathcal{F}^{(1)} \wedge C_2^{(0)} + \mathcal{F}^{(0)} \wedge C_2^{(1)}] \\
  \label{eq:2ndorderbosons}
  S_{(2)}^{(b)} &= -\frac{T_0^2}{2\pi g_s} \int d^{4}\xi \sqrt{-|M_0|}\left[\frac{1}{2}\mbox{Tr}(M_0^{-1} M_0^{(2)})\right] \nonumber \\
  &  \qquad \qquad -\frac{T_0^2}{2\pi g_s} \int d^{4}\xi \sqrt{-|M_0|}\left[\frac{1}{8}[\mbox{Tr}(M_0^{-1} M_0^{(1)})]^2 -\frac{1}{4} \mbox{Tr}(M_0^{-1}M_0^{(1)} M_0^{-1}M_0^{(1)}) \right] \nonumber\\
             &~~~~~~\qquad \qquad + \frac{T_0^2}{2\pi} \int[\mathcal{F}^{(0)} \wedge C_2^{(2)} + \mathcal{F}^{(1)} \wedge C_2^{(1)} + \mathcal{F}^{(2)} \wedge C_2^{(0)}].
\end{align}

\noindent The lowest order piece, $S_{(0)}^{(b)}$, is that which was calculated in section~\ref{tension}.

\subsection{Bosonic Fluctuations}
Continuing with the analysis, we will first show that the first order bosonic action, $S_{(1)}^{(b)}$, vanishes, up to total derivatives, confirming our previous result that we are fluctuating around a classical solution.  Next, we will find the Hamiltonian eigenvalues, $\omega$, for the quadratic fluctuations and use them to calculate the one-loop correction to the $k$-string free energy.

\subsubsection{First Order Bosonic Action}\label{1storderbosons}

Evaluating $S_{(1)}^{(b)}$ at the classical solution, $\psi = \psi_0,$  we find that it vanishes up to total derivative terms:
\begin{align}
    S_{(1)}^{(b)} &= \int d^4\xi\left[\frac{(k-M/2) \sin\theta}{4\pi}\delta F_{tx} - \frac{M}{8 \pi^2} F_{tx}^{(0)}~(\partial_{\phi}\delta\theta_m - \partial_{\theta}\delta\phi_p)\right].
  \end{align}

\noindent This confirms our earlier result that we are fluctuating around a classical solution.
\subsubsection{Second Order Bosonic Action}

The second order bosonic action, after some simplifications, takes the covariant form:
\begin{align}\label{eq:2ndorderaction}
  S_{DBI_3}^{(b)} = -\int d^4\xi\sqrt{-|g^{(\mbox{eff})}|}&\Big{\{}c_X\sum_{i=2,3}[\nabla^{\mu} \delta X^i\nabla_{\mu} \delta X^i]  + c_A [\frac{1}{16\pi}\delta F^{\mu\nu}\delta F_{\mu\nu} + \delta A_{\mu}j^{\mu}] + \nonumber\\ \qquad &+c_{\tau}[\nabla^{\mu}\delta\tau\nabla_{\mu}\delta\tau + m_{\tau}^2\delta\tau^2 + \nabla^{\mu}\Psi\nabla_{\mu}\Psi - R\Psi^2] \\ & \qquad \qquad + \mbox{Total Derivatives}\nonumber\Big{\}},
\end{align}

\noindent where the covariant derivative, $\nabla_{\mu}$, is with respect to an effective metric, $g^{(\mbox{eff})}$, on the D3-brane
\begin{align}\label{eq:geff}
 ds^2 &= g^{(\mbox{eff})}_{\mu\nu}d\xi^{\mu}d\xi^{\nu} = g_{xx}(-dt^2 + dx^2) + \frac{2}{R}(d\theta^2 + \sin^2\theta d\phi^2).
\end{align}

\noindent This effective metric has the same topology, $R^{1,1} \times S^2$, and scalar curvature, $R$, as the induced metric~\ref{eq:ginducedsimplified}.  The field $\Psi$ is a combination of the fields $\delta\psi$, and $\delta\phi_p$
\begin{align}\label{eq:Psi}
   \Psi \equiv \delta\psi + 2\cos\theta \delta\phi_p,
\end{align}
and contains all the contributions of $\delta\psi$ and $\delta\phi$ to the quadratic action suggesting a redundancy in the fields. We discuss this below.
\noindent The covariantly conserved $U(1)$ gauge current is given by
\begin{align}
    j^{\mu} &= (-Q_{\Psi} \nabla_x \Psi, ~Q_{\Psi} \nabla_t \Psi,~ -Q_{\tau} \csc\theta \nabla_{\phi} \delta\tau,~ Q_{\tau} \csc\theta\nabla_{\theta} \delta\tau ),
\end{align}

\noindent The various constants in the previous few equations are
\begin{align}\label{eq:constants}
    g_{xx} &= \frac{1}{d \sqrt{h_0}},~~~d = 1 +  b^2\tan^2{\frac{\psi_0}{2}},~~~c_A = \frac{2\sqrt{d}}{g_s},~~~c_x = \frac{\sqrt{d} T_0^3 \varepsilon^{4/3} K_0}{2 b g_s^2M}\nonumber\\
    c_\tau &= \frac{b \sqrt{d} M T_0}{32 \pi^2},~~~ m_{\tau}^2 = \frac{b^2(31 - 4\cos\psi_0) + 10\cos^2(\psi_0/2)}{45b^2}R\nonumber\\
    Q_{\tau} &=\frac{T_0^2}{6b^2 g_s M}\sec^2{\frac{\psi_0}{2}},~~~Q_{\Psi} = \frac{d^{3/2} g_s M}{8\pi^2K_0 \varepsilon^{4/3}},~~~ R = \frac{2 \sec^2 \frac{\psi}{2}}{h_0^{1/2} \varepsilon^{4/3} K_0} .
\end{align}

The Euler-Lagrange equations for the boson fields, derived from the action~\ref{eq:2ndorderaction}, take the form:
\begin{align}\label{eq:Xeqm}
   &\nabla^2 \delta X^i = 0, i = 2,3\\
   \label{eq:taueqm}
   &\nabla^2\delta\tau - m_{\tau}^2\delta\tau + \frac{c_A}{2 c_{\tau}}Q_{\tau}  \csc\theta \delta F_{\theta\phi} = 0 \\
   \label{eq:Psieqm}
   &\nabla^2\Psi + R\Psi + \frac{c_A}{2c_{\tau}} Q_{\Psi} \delta F_{tx} = 0 \\
   \label{eq:Aeqm}
   &\nabla^{\mu}\delta F_{\mu\nu} - 4\pi j_{\nu} = 0.
\end{align}

Observe that we found no field equation for $\delta \theta_m$ and that a field redefinition absorbs $\delta \phi_p$ in $\Psi$.  This is consistent with the way we arrived at the D3 brane through the D5 brane of the K-S background via a deformed conifold where the base is an $S^3\times S^2$ and the diffeomorphism gauge is fixed. The $\tau \rightarrow 0$ limit shrinks the $S^2$ and yields $M$ fractional D3 branes.  From this point of view, $\theta_m$ and $\phi_p$ were already fixed  and the absence of any fluctuations of these fields is equivalent to there being no residual gauge freedom in fixing the coordinates.  One might wonder whether the absence of field equations for $\theta_m$ and $\phi_p$ could be related to a degenerate coordinate choice.  Indeed by following ~\cite{Bigazzi}, and recalculating the Lagrangian after applying the following coordinate transformation
\begin{align}
  W \equiv \theta_m \cos\phi_p \nonumber\\
  Z \equiv \theta_m \sin\phi_p
\end{align}
we again find no field equations for $\delta W$ or $\delta Z$, up to  total derivatives.

\subsection{Bosonic Eigenvalues}
We now set out to solve the bosonic equations~\ref{eq:Xeqm}~-~\ref{eq:Aeqm}.
Notice that the composite field $\Psi$ looks like a tachyon with an electric source $\delta F_{tx}$.  With the definition of the Riemann curvature tensor
\begin{equation}\label{eq:Riemann}
  ?R^\alpha_\beta\mu\nu? \delta A^\beta = [\nabla_{\mu}, \nabla_{\nu}] \delta A^{\alpha},
\end{equation}

\noindent we can cast the $U(1)$ gauge field equations~\ref{eq:Aeqm} into the following form:
\begin{align}\label{eq:Aeqmsimplified}
4\pi j_{\nu} &= \nabla^{\mu}\delta F_{\mu\nu} = \nabla^{\mu}(\nabla_{\mu} \delta A_{\nu} - \nabla_{\nu} \delta A_{\mu}) \nonumber\\
                       &= \nabla^{\mu}\nabla_{\mu}\delta A_{\nu} - \nabla_{\nu} \nabla_{\mu} \delta A^{\mu} - ?R^\mu_\alpha\mu\nu?\delta A^{\alpha} \nonumber\\
                       &= \nabla^{\mu}\nabla_{\mu}\delta A_{\nu} - \nabla_{\nu} \nabla_{\mu} \delta A^{\mu} - R_{\alpha\nu}\delta A^{\alpha}.
\end{align}
We can further simplify these three equations by noticing that the Ricci tensor has only two non-vanishing components:
\begin{align}\label{eq:RicciComponents}
R_{\theta\theta} &= 1,~~~R_{\phi\phi} = \sin^2(\theta),~~~\mbox{all others zero}.
\end{align}

Working in the temporal gauge, $\delta A^{t} = 0$, the Gauss' law constraint is identified in equation~\ref{eq:Aeqmsimplified} as
\begin{align}\label{eq:U1constraint1}
          \nabla_t \nabla_{\mu} \delta A^{\mu} = -4\pi g_{xx} Q_{\Psi} \nabla_x \Psi.
\end{align}

We try the following ansatz for $\delta A_{\mu}$, $\Psi$, and $\delta\tau$:
\begin{align}
   \label{eq:solA}
   \delta A_i &= \int dp~d\omega \sum_{l=0}^{\infty}\sum_{m=-l}^{m=l} \tilde{A_i}^{(lm)} (p,\omega)~e^{i(px - \omega t)}~Y_i^{(lm)}(\theta,\phi),~~i = x,\theta,\phi  \\
   \label{eq:solPsi}
   \Psi &= \int dp~d\omega \sum_{l=0}^{\infty}\sum_{m=-l}^{m=l} \tilde{\Psi}^{(lm)}(p,\omega)~e^{i(px - \omega t)}~Y_{(lm)}(\theta,\phi) \\
   \label{eq:soltau}
   \delta\tau &= \int dp~d\omega \sum_{l=0}^{\infty}\sum_{m=-l}^{m=l} \tilde{\tau}^{(lm)} (p,\omega)~e^{i(px - \omega t)}~Y_{(lm)}(\theta,\phi),
\end{align}

\noindent where the $Y_i^{(lm)}(\theta,\phi)$ are
\begin{align}\label{eq:vectorYlm}
Y_x^{(lm)} &\equiv Y_{(lm)}(\theta,\phi)\nonumber\\
Y_{\theta}^{(lm)} &\equiv \frac{\csc{\theta}}{\sqrt{l(l+1)}}\partial_{\phi} Y_{(lm)}(\theta,\phi)\nonumber\\
Y_{\phi}^{(lm)} &\equiv \frac{-\sin{\theta}}{\sqrt{l(l+1)}}\partial_{\theta} Y_{(lm)}(\theta,\phi),
\end{align}

\noindent $Y_{\theta}^{(lm)}$ and $Y_{\phi}^{(lm)}$ are \emph{vector} spherical harmonics which satisfy the eigenvalue equation
\begin{align}\label{eq:vectorYlmeigen}
   \hat{L}^2 Y_j^{(lm)} &= [-l(l+1) + 1] Y_j^{(lm)}, j=\theta,\phi
\end{align}

\noindent where
\begin{align}
   \hat{L}^2 &= \frac{1}{\sin{\theta}}\partial_{\theta}\sin{\theta}\partial_{\theta} + \frac{1}{\sin^2{\theta}}\partial_{\phi}^2.
\end{align}

Using an ansatz with $\tilde{A_{\theta}} = \tilde{A_{\phi}}$, the Gauss law constraint~\ref{eq:U1constraint1} becomes simply
\begin{align}\label{eq:U1constraint2}
    \nabla_x \nabla_t\delta A_x = -4\pi g_{xx}^2 Q_{\Psi} \nabla_x \Psi.
\end{align}

This can be used to simplify the $\delta A_x$ equation to the non-dynamical form
\begin{align}
   \hat{L}^2 \tilde{A_x} = 0
\end{align}

\noindent which means $l=0$ for the coupled fields $\delta A_x$ and $\Psi$.  The $\Psi$ equation~\ref{eq:Psieqm} then becomes the eigenvalue equation
\begin{align}\label{eq:beigenproblem1}
   [\omega^2 - p^2 - m_{\Psi}^2] \tilde{\Psi} = 0
\end{align}
where $m_{\Psi}^2$ is positive and is,
\begin{equation}\label{eq:Psimass}
   m_{\Psi}^2 = 4\pi g_{xx}^2 Q_{\Psi}^2 \frac{c_A}{2 c_{\tau}} - R
               = \frac{1 + (1-b^2)\cos\psi_0}{b^2}R   \end{equation}
  where
  \[17.5055 \frac{T_0}{g_s M} \le m_{\Psi}^2 < \infty.\label{eq:Psimassrange}
\]

Next, we have the coupled fields $\delta A_{\theta}$, $\delta A_{\phi}$, and $\delta\tau$, whose eigenvalue problem becomes
\begin{align}\label{eq:beigenproblem2}
&[\omega^2 - p^2 - g_{xx}\frac{R}{2}l(l+1)] \tilde{A_i} - \mathcal{H}_A\tilde{\tau} = 0,~~~i=\theta,\phi \nonumber\\
&[\omega^2 - p^2 - g_{xx}\frac{R}{2}l(l+1) - g_{xx}m_{\tau}^2]\tilde{\tau} - \mathcal{H}_{\tau}\tilde{A_{\theta}} = 0,\nonumber\\
  \mathcal{H}_{\tau} &= -g_{xx}Q_{\tau} \frac{c_A}{2 c_\tau}\sqrt{l(l+1)},~~~\mathcal{H}_A = -\frac{8\pi}{R}g_{xx}Q_{\tau}\sqrt{l(l+1)}.
\end{align}

Finally, the two massless equations~\ref{eq:Xeqm} can be solved with
\begin{equation}
   \delta X^i = \int dp~d\omega~\tilde{X^i}_{(lm)}(p,\omega) e^{i(p x - \omega t)}Y_{(lm)}(\theta,\phi)~~~i=2,3
\end{equation}

\noindent yielding two identical eigenvalue problems
\begin{align}\label{eq:beigenproblem3}
   [\omega^2 - p^2 - g_{xx}\frac{R}{2}l(l+1)]\tilde{X^i} = 0,~~~i=2,3.
\end{align}

We now organize equations~\ref{eq:beigenproblem1},~\ref{eq:beigenproblem2}, and~\ref{eq:beigenproblem3} into the succinct form:
\begin{align}\label{eq:beigenproblem}
  \omega^2 \left(\begin{array}{l}
             \tilde{\Psi}\\
             \tilde{X^2}\\
             \tilde{X^3}\\
             \tilde{\tau}\\
             \tilde{A_{\theta}} \\
             \tilde{A_{\phi}}
          \end{array}
   \right) &= \mathcal{H}_{(b)}^2
   \left(\begin{array}{l}
             \tilde{\Psi}\\
             \tilde{X^2}\\
             \tilde{X^3}\\
             \tilde{\tau}\\
             \tilde{A_{\theta}}\\
             \tilde{A_{\phi}}
          \end{array}
   \right)
\end{align}

\noindent where the square of the bosonic Hamiltonian, $\mathcal{H}_{(b)}^2$, is given by
\begin{align}\label{eq:bHamiltoniansquared}
   \mathcal{H}_{(b)}^2 &= \left(
                      \begin{array}{l l l l l l}
                      \omega_1^2 & 0 & 0 & 0 & 0 & 0\\
                      0 & \omega_2^2 & 0 & 0 & 0 & 0\\
                      0 & 0 & \omega_2^2 & 0 & 0 & 0\\
                      0 & 0 & 0 & \omega_2^2 + g_{xx}m_{\tau}^2 & \mathcal{H}_{\tau} & 0\\
                      0 & 0 & 0 & \mathcal{H}_A & \omega_2^2 & 0 \\
                      0 & 0 & 0 & \mathcal{H}_A & 0 & \omega_2^2
                      \end{array}
                      \right)
\end{align}

\noindent where
\[
  \omega_1^2 = p^2 + g_{xx}m_{\Psi}^2 \qquad \mbox{and} \qquad
  \omega_2^2 = p^2 + g_{xx}\frac{R}{2}l(l+1).
\]

\noindent  Furthermore, the six eigenvalues of this matrix are the squares of the bosonic energy eigenvalues, $\omega$.
\begin{align}\label{eq:bosoniceigenvalues}
  \omega^2 &= \left\{ \begin{array}{l}
                       \omega_1^2 \\
                       \omega_2^2~\mbox{3-fold degenerate} \\
                       \omega_{\pm}^2
                       \end{array}
              \right.
\end{align}

\noindent where
\begin{align}
  \omega_{\pm}^2 &= \omega_2^2 + g_{xx}\frac{m_{\tau}^2}{2}\left(1 \pm \sqrt{1+ \frac{4 \mathcal{H}_A \mathcal{H}_{\tau}}{g_{xx}^2 m_{\tau}^4}}\right).
\end{align}

\subsection{Fermionic Eigenvalues}\label{fermioniceigenvalues}
Here we present a detailed solution to the fermionic field equations.
We start by variation of the fermionic action, equation~\ref{eq:Sfexpanded}, where $\delta\Theta$ are the fermions fluctuated about the zero field.  The D-brane Dirac equation becomes,
\begin{align}\label{eq:Diraceq}
  [\left(M_0^{-1}\right)^{\alpha\beta}\Gamma _{\alpha}\partial_{\beta}+M_1+M_2+M_3]\delta\Theta_1
\end{align}

\noindent where $M_0$ is as before, and $M_1$, $M_2$, $M_3$, and the pulled back $\Gamma_{\alpha}$ matrices are found in appendix~\ref{app:fermions}.
There it is shown that $M_1$ contains the spin connection, $?\Omega_{\beta}^{\bar{a}\bar{b}}?$, which is antisymmetric in $\bar{a}$ and $\bar{b}$, and has the following non-vanishing components in the $\tau \to 0$ limit:
\begin{align}\label{eq:spinconnection}
  ?\Omega_{\theta}^{\bar{4}\bar{9}}? &= ?\Omega_{\theta}^{\bar{8}\bar{7}}? = \frac{\sin\psi_0}{2},~~~?\Omega_{\theta}^{\bar{5}\bar{9}}? = ?\Omega_{\theta}^{\bar{6}\bar{8}}? =  \sin^2\frac{\psi_0}{2} \nonumber\\
  ?\Omega_{\phi}^{\bar{5}\bar{4}}? &= ?\Omega_{\phi}^{\bar{7}\bar{6}}? = \cos\theta,~~~?\Omega_{\phi}^{\bar{9}\bar{4}}? = ?\Omega_{\phi}^{\bar{7}\bar{8}}? = \sin\theta~\sin^2\frac{\psi_0}{2} \nonumber\\
  ?\Omega_{\phi}^{\bar{5}\bar{9}}? &= ?\Omega_{\phi}^{\bar{6}\bar{8}}? = \frac{1}{2}\sin\theta~\sin\psi_0.
\end{align}

This spin connection is quite distinct from the spin connection  that one gets from the 4-D spin connection inherited on the D3-brane through the induced metric.  That spin connection has just one non-vanishing component:
\begin{align}
  (\Omega^{(4d)}?)_\phi^{\bar{3}\bar{2}}? &= \cos\theta.
\end{align}

\noindent This complication stems from the fact that the 10-d KS background has paired $\theta$ and $\phi$ coordinates, $\theta_1, \theta_2, \phi_1$, and $\phi_2$, and is the main source of complexity in the fermionic field equations.

We proceed by using a harmonic ansatz for $\delta\Theta$ in the Dirac equation~\ref{eq:Diraceq} \cite{ingo}:
\begin{align}\label{eq:Thetasoln}
\delta\Theta &= \int dp~d\omega\sum_{l}e^{i(p x - \omega t)}\tilde{\Theta}_{lm}(p,\omega) \circ \Phi_{lm}(\theta,\phi)
\end{align}

\noindent where $\Phi_{lm}(\theta,\phi)$ is a 32 component complex spinor, whose components are arbitrary functions of $\theta$ and $\phi$, and $\tilde{\Theta}_{lm}(p,\omega)$ is a 32 component spinor of Grassman valued expansion coefficients.  The \emph{component} product $\circ$ is a commutative operator, defined for N component vectors or spinors as:
\begin{align}\label{eq:compproduct}
   A\circ B &= \left(\begin{array}{l} A_1 \\ A_2 \\ \vdots \\ A_N  \end{array} \right)
                      \circ
               \left(\begin{array}{l} B_1 \\ B_2 \\ \vdots \\ B_N \end{array}\right)
                      \equiv
               \left(\begin{array}{l} A_1 B_1 \\ A_2 B_2 \\ \vdots \\ A_N B_N \end{array} \right).
\end{align}

With solution~\ref{eq:Thetasoln} for $\delta\Theta$, the Dirac equation~\ref{eq:Diraceq}, can be reorganized and expressed as the eigenvalue problem
\begin{align}
   \omega \tilde{\Theta}\circ \Phi = \mathcal{H}_{(f)} \tilde{\Theta}\circ \Phi
\end{align}

\noindent where the Hamiltonian $\mathcal{H}_{(f)}$ has the block diagonal form
\begin{align}
    \mathcal{H}_{(f)} &= \left(\begin{array}{l l l l}
                                   \mathcal{H}^{(f)}_1 & 0 & 0 & 0 \\
                                   0 & \mathcal{H}^{(f)}_2 & 0 & 0 \\
                                   0 & 0 & \mathcal{H}^{(f)}_1 & 0 \\
                                   0 & 0 & 0 & \mathcal{H}^{(f)}_2
                                \end{array}
                         \right).
\end{align}

\noindent The exact forms of the eight by eight matrices $\mathcal{H}^{(f)}_1$ and $\mathcal{H}^{(f)}_2$, as well as a detailed calculation of their eigenvalues, are found in appendix~\ref{app:Hfermions}.  There they are found to share the same eight eigenvalues
\begin{align}\label{eq:fermioniceigenvalues}
   \omega &= \left\{\begin{array}{l}
                    \pm\sqrt{c_{10}(p,l) + \sqrt{c_8(p,l)} \pm \sqrt{c_{9+}(p,l)}} \\
                    \pm\sqrt{c_{10}(p,l) - \sqrt{c_8(p,l)} \pm \sqrt{c_{9-}(p,l)}}
                    \end{array}
                    \right.
\end{align}

\noindent where $c_8(p,l), c_9(p,l)$, and $c_{10}(p,l)$ are given in appendix~\ref{app:Hfermions}.
\subsection{One Loop Energy Corrections: The Luscher Term}
Following closely~\cite{pandoz,Bigazzi}, we now demonstrate our semi-classical approach to calculating the one loop energy corrections of k-strings.  We calculate the one loop energy corrections as the sum of the bosonic and fermionic one loop energies
\begin{align}\label{eq:oneloopEnergy}
  E_1 &=  E_1^{(b)} + E_1^{(f)}
\end{align}

\noindent where
\begin{align}
   E_1^{(b)} &= \sum_{bosons} \omega_c^{(b)} \\
   E_1^{(f)} &= -\sum_{fermions} \omega_c^{(f)}
\end{align}

\noindent where $\omega_c^{(b)}$ and $\omega_c^{(f)}$ are the positive, classical energy eigenvalues of the bosons and fermions in equations~\ref{eq:bosoniceigenvalues} and~\ref{eq:fermioniceigenvalues}, respectively.

The one loop bosonic energy is given by
\begin{align}
  E_1^{(b)} =  \sum_p\omega_1 + 3\sum_{p,l,m}\omega_2 + \sum_{p,l,m}\omega_+ + \sum_{p,l,m} \omega_-.
\end{align}

\noindent As shown in appendix~\ref{app:regularization}, only the massless modes contribute to the $1/L$ piece and we find that
\begin{align}
  E_1^{(b)} &= V_{\mbox{\small{L\"uscher}}} + V_c(k).
\end{align}

\noindent where we interpret $1/L$ dependent $V_{\mbox{\small{L\"uscher}}}$ as a L\"uscher term
\begin{align}
V_{\mbox{\small{L\"uscher}}} = -\frac{\pi}{3L},
\end{align}

\noindent and the term $V_c(k)$ is constant in $L$ but dependent on $k$.  The L\"uscher term represents a Coulomb-like potential for largely separated quarks and only depends on the massless fields and therefore is $k$ independent. Here we have four massless modes, two coming from the bosonic fields $X^2$ and $X^3$, and two modes from the photon, see Eqs.[\ref{eq:pmsum},\ref{eq:2sum}].  This is to be compared to the L\"uscher terms for a confining four dimensional gauge theory were there are two massless modes and to the gravity theory where there are two massless modes coming from quantizing the string \cite{maldecena_email}. Furthermore the L\"uscher term (since it is constant in $k$) as well as  $V_c(k)$ respect the symmetry for $k \to M - k$ and this is shown in appendix~\ref{app:regularization}.
As it turns out, regulating the fermionic eigenvalues is quite difficult (see \ref{eq:fermioniceigenvalues}) but as there are no massless modes in this sector, we do not expect the fermions to contribute to the Luscher term.  We should remark that other researchers have found $1/L$ contributions at the {\em classical}  level in the case of a Maldecena-Nunez supergravity background \cite{hartnoll}.  

\section{Conclusions}
One of the hopes of the gauge/gravity correspondence is that we may soon have enough machinery to make predictions about present day QCD even though the duality is coached around supergravity and supersymmetric gauge theories.  In this work we focused on the Klebanov-Strassler supergravity background.
In order to probe this background, we calculated the field equations for the D3-brane action and found that, in agreement with Herzog and Klebanov, a near $\sin(\pi k/M) $ behavior for the string tension of the $k$-strings.  We went further by  calculating the one loop corrections to the energy through quadratic fluctuations about the classical configuration. Here we included both the bosonic and fermionic fluctuations in the analysis and found explicit formulas for their frequencies. It is interesting to note that the number of bosonic fields that contribute to the quadratic fluctuations is more akin to fluctuations on a D5 brane.   This suggests that the limiting procedure to arrive at the D3 brane by going to the tip of a D5 brane preserves the diffeomorphism symmetry of the D5 brane thus eliminating the fluctuations of two more of the bosonic fields.   By using the massless modes of the quadratic action we were able to find a $1/L$ correction to the $k$-string energy which we interpret as a  L\"uscher term. This result is robust since the massive modes contribute terms proportional to $\exp(-m L)$ which has no $1/L$ contribution.  For the $2+1$ dimensional case, lattice calculations are already available for the spectrum of $k$-strings, including $L$ dependence, for certain values of $N$ \cite{Athenodorou:2007du,Athenodorou:2008qg}.  These lattice results also suggests that the closed string spectrum is in the same universality class of the Nambu-Goto string. 

There are a number of questions that our work naturally suggests. First is whether the  $k$-string excitations as computed from the supergravity backgrounds are also among a universality class or do they depend explicitly on the model.  One way to understand this is to continue this line of research using other supergravity backgrounds such as the Maldecena-N\`u\~{n}ez \cite{mn} and Witten QCD backgrounds.  At the classical level they all look quite similar, however we expect the excitations to clearly determine the `universal' modes from the model-dependent modes.  This work is presently underway.

\section*{Acknowledgments}
The authors would like to gratefully acknowledge conversations with A. Armoni, B. Bringolz, S. Hartnoll, I. Kirsch, V.P. Nair, A. Ramallo, and M. Teper.
We would also like to thank M. Hansen, J. Samorajski, and W. Merrell for collaboration in the early stages of this project.
This work is  partially supported by Department of Energy under
grant DE-FG02-95ER40899 to the University of Michigan and the National Science Foundation under award PHY  - 0652983 to the University of Iowa.

\appendix

\section{Fermionic Action Definition and Results}
\label{app:fermions}

The general form of the fermionic portion of the D3-brane action used in this paper was found in~\cite{m4}.
\begin{align}\label{eq:Sf}
    S^{(F)}_{DBI_p}=\frac{\mu_p}{2 g_s}\int d^{p+1}\xi e^{-\Phi}\sqrt{-\det(M_0)}~&\overline{\Theta}\{\left(M_0^{-1}\right)^{\alpha\beta}\Gamma _{\alpha}D_{\beta}^{(0)} - \Delta^{(1)} \nonumber\\        &-\overset{\vee}{\Gamma}_{D_p}^{-1}(M_0^{-1})^{\alpha\beta}\Gamma_{\beta}W_{\alpha} + \overset{\vee}{\Gamma}_{D_p}^{-1}\Delta^{(2)}\}\Theta
\end{align}

\noindent where
\begin{align}\label{eq:Sfpieces}
M_0 &= g + \mathcal{F} \nonumber\\
D_{\alpha}^{(0)} &= \partial_{\alpha} + \frac{1}{4}?\Omega_\alpha^{\bar{a}\bar{b}}?\Gamma_{\bar{a}\bar{b}} + \frac{1}{4 \cdot 2!} H_{\alpha np}\Gamma^{np} \nonumber\\
W_{\alpha} &= \frac{1}{8}\left[ F_n\Gamma^n + \frac{1}{3!}(F_{npq} + C_0 H_{npq})\Gamma^{np}+ \frac{1}{2\cdot5!}(F_{npqrs} + H_{\lbrack npq}C_{rs\rbrack})\Gamma^{npqrs}\right]\Gamma_{\alpha} \nonumber\\
\Delta^{(1)} &= \frac{1}{2}(\Gamma^m\partial_m\Phi + \frac{1}{2\cdot3!} H_{mnp}\Gamma^{mnp}) \nonumber\\
\Delta^{(2)} &= -\frac{1}{2}e^{\Phi}[F_m \Gamma^m + \frac{1}{2\cdot3!}(F_{mnp} + C_0 H_{mnp})\Gamma^{mnp}] \nonumber\\
\overset{\vee}{\Gamma}_{D_p} &= (-1)^{\frac{(p-2)(p-3)}{2}} \frac{\varepsilon^{\alpha_1 \dots \alpha_{p+1}}\Gamma_{\alpha_1\dots\alpha_{p+1}}}{(p+1)!\sqrt{-\det M_0}}\sum_q \frac{\Gamma^{\lbrack\beta_1 \dots \beta_{2q}\rbrack}}{q!2^q} \mathcal{F}_{\beta_1\beta_2}\cdots\mathcal{F}_{\beta_{2q-1}\beta_{2q}}
\end{align}
\noindent and where $\Gamma^{abc\dots}$ = $\Gamma^a\Gamma^b\Gamma^c\dots$.
Our index convention is that latin indices, ($a,b,c,\dots$), are 10-d Bosonic supergravity indices, and greek indices, ($\mu,\nu,\alpha,\dots$), Dp-brane world volume indices.  Furthermore, latin indices with an overbar, ($\bar{a},\bar{b},\bar{c},\dots$), are flat space-time indices.  The 10-d flat $\Gamma^{\bar{a}}$ matrices satisfy a Clifford algebra:
\begin{align}\label{eq:10DflatClifford}
\left\{ \Gamma^{\bar{a}}, \Gamma^{\bar{b}} \right\} = 2\eta^{\bar{a}\bar{b}}
\end{align}

We use the following representation for the 10-d flat $\Gamma^{\bar{a}}$ matrices:
\begin{align}\label{eq:flatgamma}
  \Gamma^{\bar{0}} &= -i \sigma^1 \otimes \sigma^1 \otimes \sigma^1 \otimes \sigma^1 \otimes \sigma^3,~~~\Gamma^{\bar{1}} = \sigma^1 \otimes \sigma^1 \otimes \sigma^1 \otimes \sigma^2 \otimes \sigma^0 \nonumber\\
  \Gamma^{\bar{2}} &= \sigma^1 \otimes \sigma^3 \otimes \sigma^0 \otimes \sigma^0 \otimes \sigma^0,~~~\Gamma^{\bar{3}} = \sigma^2 \otimes \sigma^0 \otimes \sigma^0 \otimes \sigma^0 \otimes \sigma^0 \nonumber\\
  \Gamma^{\bar{4}} &= \sigma^1 \otimes \sigma^1 \otimes \sigma^3 \otimes \sigma^0 \otimes \sigma^0,~~~\Gamma^{\bar{5}} = \sigma^1 \otimes \sigma^2 \otimes \sigma^0 \otimes \sigma^0 \otimes \sigma^0 \nonumber\\
  \Gamma^{\bar{6}} &= -\sigma^1 \otimes \sigma^1 \otimes \sigma^1 \otimes \sigma^1 \otimes \sigma^1,~~~\Gamma^{\bar{7}} = \sigma^1 \otimes \sigma^1 \otimes \sigma^1 \otimes \sigma^1 \otimes \sigma^2 \nonumber\\
  \Gamma^{\bar{8}} &= \sigma^1 \otimes \sigma^1 \otimes \sigma^2 \otimes \sigma^0 \otimes \sigma^0,~~~\Gamma^{\bar{9}} = \sigma^1 \otimes \sigma^1 \otimes \sigma^1 \otimes \sigma^3 \otimes \sigma^0
\end{align}

\noindent where $\otimes$ means tensor product, and the $\sigma^{\mu}$ are the Pauli spin matrices, augmented with the identity:
\begin{align}\label{eq:Paulispin}
   \sigma^0 &= \left(\begin{array}{l l}
                     1 & 0 \\
                     0 & 1
                     \end{array}
                \right),~~~
   \sigma^1 &= \left(\begin{array}{l l}
                       0 & 1 \\
                       1 & 0
                      \end{array}
                \right),~~~
   \sigma^2 &= \left(\begin{array}{l l}
                      0 & -i \\
                      i & 0
                     \end{array}
               \right),~~~
   \sigma^3 &= \left(\begin{array}{l l}
                      1 & 0 \\
                      0 & -1
                     \end{array}
               \right)
\end{align}

The 10-d curved $\Gamma^{a}$ matrices are related to the 10-d flat $\Gamma^{\bar{a}}$ matrices via the viel-biens
\begin{align}\label{eq:10dGammaRelation}
\Gamma^a = ?e_{\bar{a}}^a? \Gamma^{\bar{a}},~~~\Gamma_{a} = ?e_a^{\bar{a}}?\Gamma_{\bar{a}}
\end{align}

\noindent and so the 10-d curved $\Gamma^a$ matrices will,in general, satisfy a \emph{curved} Clifford algebra:
\begin{align}\label{eq:10dcurvedClifford}
\left\{ \Gamma^a, \Gamma^{b} \right\} = 2 ?e_{\bar{a}}^a??e_{\bar{b}}^b?\eta^{\bar{a}\bar{b}} = 2 G^{ab}
\end{align}.

Also, $\Gamma^{\alpha}$ are the 10-curved $\Gamma^a$ matrices pulled back onto the D3-brane:
\begin{align}\label{eq:D3braneGamma}
\Gamma_{\alpha} &= \frac{\partial X^a}{\partial \xi^{\alpha}} \Gamma_a \nonumber\\
\Gamma^{\alpha} &= g^{\alpha\beta}\Gamma_{\beta}
\end{align}

The Spin Connection, $?\Omega_{\alpha}^{\bar{a}\bar{b}}?$ in \ref{eq:Sfpieces} is the pullback of the full 10-D spin connection, $?\Omega_a^{\bar{a}\bar{b}}?$ (only the first index is pulled back to the D3-brane):
\begin{align}\label{eq:10dSpinConnection}
  ?\Omega_a^{\bar{a}\bar{b}}? &= \frac{1}{2} ?e_a^{\bar{c}}? ( ?\eta_{\bar{d}\bar{c}}?  ?\eta^{\bar{e}\bar{a}}? ?\eta^{\bar{f}\bar{b}}?  - ?\delta_{\bar{d}}^{\bar{a}}? ?\eta^{\bar{e}\bar{b}}? ?\delta_{\bar{c}}^{\bar{f}}? - ?\delta_{\bar{d}}^{\bar{b}}? ?\delta_{\bar{c}}^{\bar{e}}? ?\eta^{\bar{f}\bar{a}}? )?C^{\bar{d}}_{\bar{e}\bar{f}}? \nonumber\\
  ?C^{\bar{a}}_{\bar{b}\bar{c}}? &= (?e_{\bar{b}}^a? ?e_{\bar{c}}^b? - ?e_{\bar{b}}^b? ?e_{\bar{c}}^a?)\partial_b ?e_a^{\bar{a}}?
\end{align}

For the KS background at $\tau = 0$, the definitions~\ref{eq:Sfpieces} simplify to
\begin{align}\label{eq:simplifiedSfpieces}
  D_{\alpha}^{(0)} &= \partial_{\alpha} + \frac{1}{4} ?\Omega_{\alpha}^{\bar{a}\bar{b}}?\Gamma_{\bar{a}\bar{b}} \nonumber\\
  W_{\alpha} &= \frac{1}{8 \cdot 3!} F_{npq}\Gamma^{npq} \Gamma_{\alpha} \nonumber\\
  \Delta^{(1)} &= 0,~~~~~ \Delta^{(2)} = -\frac{1}{4\cdot 3!} F_{mnp}\Gamma^{mnp} \nonumber\\
  \overset{\vee}{\Gamma}_{D_3} &= \frac{\varepsilon^{\alpha\beta\mu\nu} \Gamma_{\alpha\beta\mu\nu}}{2! 4! \sqrt{-\det M_0}} \Gamma^{\lambda\rho}\mathcal{F}_{\lambda\rho}
\end{align}

\noindent whereupon plugging these into the fermionic action~\ref{eq:Sf} for D3-branes gives:
\begin{align}
  S_{D_3}^{(F)} &= \frac{T_0^2}{4\pi g_s} \int d^4\xi \sqrt{-\det M_0} \overline{\Theta} [(M_0^{-1})^{\alpha\beta}\Gamma_{\alpha}\partial_{\beta} + M_1 + M_2 + M_3 ] \Theta \nonumber\\
  M_1 &= \frac{1}{4} (M_0^{-1})^{\alpha\beta}\Gamma_{\alpha}?\Omega_{\beta}^{\bar{a}\bar{b}}?\Gamma_{\bar{a}\bar{b}} \nonumber\\
  M_2 &= -\frac{1}{8 \cdot 3!}\overset{\vee}{\Gamma}_{D_3}^{-1} (M_0^{-1})^{\alpha\beta}\Gamma_{\beta} F_{mnp}\Gamma^{mnp} \Gamma_{\alpha} \nonumber\\
  M_3 &= -\frac{1}{4 \cdot 3!}\overset{\vee}{\Gamma}_{D_3}^{-1} F_{mnp}\Gamma^{mnp}
\end{align}

For our solution of the KS background, we find the object $\overset{\vee}{\Gamma}_{D_3}^{-1}$ can be expressed as:
\begin{align}\label{eq:InvGammav}
  \overset{\vee}{\Gamma}_{D_3}^{-1} &= b^{-1}\cot{\frac{\psi_0}{2}}\Gamma^{\bar{6}}\Gamma^{\bar{7}}.
\end{align}

\noindent and we calculate the pulled back $\Gamma^{\mu}$ matrices to be:
\begin{align}\label{eq:Gammpb}
  \Gamma^{\mu} &= \left(\begin{array}{l}
                       h_0^{1/4}\Gamma^{\bar{0}} \\
                       h_0^{1/4}\Gamma^{\bar{1}} \\
                       \sqrt{\frac{R}{2}}(\cos{\frac{\psi_0}{2}}\Gamma^{\bar{7}} - \sin{\frac{\psi_0}{2}}\Gamma^{\bar{6}}) \\
                       -\sqrt{\frac{R}{2}}\csc{\theta}(\sin{\frac{\psi_0}{2}}\Gamma^{\bar{7}} + \cos{\frac{\psi_0}{2}}\Gamma^{\bar{6}})
                     \end{array}
                 \right).
\end{align}

\section{Fermionic Hamiltonian and Eigenvalues}
\label{app:Hfermions}

The two distinct fermionic eigenvalue equations are
\begin{align}\label{eq:feigenproblems}
   \omega \tilde{\Theta}_1\circ \Phi_1 = \mathcal{H}^{(f)}_1 \tilde{\Theta}_1 \circ \Phi_1 \\
   \omega \tilde{\Theta}_2\circ \Phi_2 = \mathcal{H}^{(f)}_2 \tilde{\Theta}_2\circ \Phi_2
\end{align}

\noindent where $\tilde{\Theta}_1\circ \Phi_1$ and $\tilde{\Theta}_2\circ \Phi_2$ are each eight component spinors.  The \emph{component} product operator, $\circ$, was defined in equation~\ref{eq:compproduct}.  The matrices $\mathcal{H}^{(f)}_i$ are
{\footnotesize \begin{align}\label{eq:Hf1}
   \mathcal{H}^{(f)}_1 &= \left(\begin{array}{l l l l l l l l}
                          -p & -c_{3+}\mathcal{O}^{(2)}_+ & c_{2+} & 0 & 0 & i c_+ & 0 & 0 \\
                          -c_{4-}\mathcal{O}^{(1)}_- & p & 0 & c_{1-} & 0 & 0 & 0 & 0 \\
                          -c_{1-} & 0 & p & c_{3-} \mathcal{O}^{(2)}_+ & 0 & 0 & 0 & i c_- \\
                          0 & -c_{2+} & c_{4+}\mathcal{O}^{(1)}_- & -p & 0 & 0 & 0 & 0 \\
                          0 & 0 & 0 & 0 & -p & -c_{3+}\mathcal{O}^{(1)}_+ & -c_{2+} & 0 \\
                          i c_- & 0 & 0 & 0 & -c_{4-}\mathcal{O}^{(2)}_- & p & 0 & -c_{1-} \\
                          0 & 0 & 0 & 0 & c_{1-} & 0 & p & c_{3-}\mathcal{O}^{(1)}_+ \\
                          0 & 0 & i c_+ & 0 & 0 & c_{2+} & c_{4+}\mathcal{O}^{(2)}_- & -p
                          \end{array}
                          \right)
\end{align}}

\noindent and
{\footnotesize \begin{align}\label{eq:Hf2}
   \mathcal{H}^{(f)}_2 &= \left(\begin{array}{l l l l l l l l}
                          -p & -c_{3+}\mathcal{O}^{(1)}_+ & c_{2+} & 0 & 0 & 0 & 0 & 0 \\
                          -c_{4-}\mathcal{O}^{(2)}_- & p & 0 & c_{1-} & -i c_- & 0 & 0 & 0 \\
                          -c_{1-} & 0 & p & c_{3-} \mathcal{O}^{(1)}_+ & 0 & 0 & 0 & 0 \\
                          0 & -c_{2+} & c_{4+}\mathcal{O}^{(2)}_- & -p & 0 & 0 & -i c_+ & 0 \\
                          0 & -i c_+ & 0 & 0 & -p & -c_{3+}\mathcal{O}^{(2)}_+ & -c_{2+} & 0 \\
                          0 & 0 & 0 & 0 & -c_{4-}\mathcal{O}^{(1)}_- & p & 0 & -c_{1-} \\
                          0 & 0 & 0 & -i c_- & c_{1-} & 0 & p & c_{3-}\mathcal{O}^{(2)}_+ \\
                          0 & 0 & 0 & 0 & 0 & c_{2+} & c_{4+}\mathcal{O}^{(1)}_- & -p
                          \end{array}
                          \right)
\end{align}}

\noindent where the constants $c_{\pm}$ and $c_{1\pm}\dots c_{4\pm}$ are
\begin{align}
  c_{1\pm} &= i \frac{c_{\pm}}{2 T^2}(3T^2 - d^{1/2}T - b_3),~~~c_{2\pm} = i \frac{c_{\pm}}{2 T^2}(3T^2 + d^{1/2}T - b_3) \nonumber\\
  c_{3\pm} &= c_{\pm}(1 - i \frac{b}{T}),~~~c_{4\pm} = c_{\pm}(1 + i \frac{b}{T})\nonumber\\
  c_{\pm} &= -\frac{ 2 b_2 T }{h_0^{1/4} d^{1/2}}(T \pm d^{1/2}) \nonumber\\
  T &= b \tan{\frac{\psi_0}{2}},~~~~d = 1 + b^2\tan^2{\frac{\psi_0}{2}} \nonumber\\
  b_2 &= \sqrt{\frac{\pi T_0}{2b^3M}},~~~b_3 = b^{-1/2} + 3b^{3/2} - 3b^2
\end{align}

\noindent and the operators $\mathcal{O}^{(i)}_{\pm}$ are
\begin{align}
  \mathcal{O}^{(1)}_{\pm} &= \partial_{\theta} \pm i \csc{\theta} \partial_{\phi}\\
  \mathcal{O}^{(2)}_{\pm} &= \cot{\theta} + \mathcal{O}^{(1)}_{\pm}
\end{align}

From inspection of the Hamiltonians~\ref{eq:Hf1} and~\ref{eq:Hf2}, and their actions on the spinors $\Phi_{ilm}$ in equations~\ref{eq:feigenproblems}, we identify the components, $\Phi^A_{ilm}(\theta,\phi),A=1\dots8,i=1,2$, of the spinors $\Phi_{ilm}(\theta,\phi)$ with three distinct functions $Y^+_{lm}(\theta,\phi)$, $Y_{lm}(\theta,\phi)$, and $Y^-_{lm}(\theta,\phi)$
\begin{align}
   \Phi_1^1 &= \Phi_1^3 = \Phi_1^6 = \Phi_1^8 = \Phi_2^2 = \Phi_2^4 = \Phi_2^5 = \Phi_2^7 = Y_{lm}(\theta,\phi) \nonumber\\
   \Phi_1^2 &= \Phi_1^4 = \Phi_2^6 = \Phi_2^8 = Y^-_{lm}(\theta,\phi),~~~\Phi_1^5 = \Phi_1^7 = \Phi_2^1 = \Phi_2^3 = Y^+_{lm}(\theta,\phi)
\end{align}

\noindent which must satisfy four coupled differential equations
\begin{align}
   \label{eq:coupled1}
   \mathcal{O}^{(1)}_{-}Y_{lm}(\theta,\phi) = \lambda_1 Y^-_{lm}(\theta,\phi) \\
   \label{eq:coupled2}
   \mathcal{O}^{(2)}_{+}Y^-_{lm}(\theta,\phi) = \lambda_2 Y_{lm}(\theta,\phi) \\
   \label{eq:coupled3}
   \mathcal{O}^{(1)}_{+}Y_{lm}(\theta,\phi) = \lambda_3 Y^+_{lm}(\theta,\phi) \\
   \label{eq:coupled4}
   \mathcal{O}^{(2)}_{-}Y^+_{lm}(\theta,\phi) = \lambda_4 Y_{lm}(\theta,\phi)
\end{align}

\noindent Eliminating $Y^-_{lm}(\theta,\phi)$ from equations~\ref{eq:coupled1} and~\ref{eq:coupled2} results in the spherical harmonic eigenvalue problem:
\begin{align}\label{eq:coupled12}
   \hat{L}^2 Y_{lm} = \lambda_1 \lambda_2 Y_{lm}
\end{align}

\noindent So we see that the $Y_{lm}(\theta,\phi)$ are indeed the spherical harmonics, as their name suggests.  Furthermore, equation~\ref{eq:coupled12} now demands that
\begin{align}
  \lambda_1 \lambda_2 &= -l(l+1)
\end{align}

\noindent Eliminating $Y^+_{lm}(\theta,\phi)$ from equations~\ref{eq:coupled3} and~\ref{eq:coupled4} results in a similar identity
\begin{align}
   \lambda_3 \lambda_4 &= -l(l+1)
\end{align}

\noindent Consistent with these two constraints, we make the following choices for the $\lambda_i$:
\begin{align}
  \lambda_1 &= \lambda_3 = 1,~~~\lambda_2 = \lambda_4 = -l(l+1)
\end{align}

\noindent and so we find $Y^+_{lm}(\theta,\phi)$ and $Y^-_{lm}(\theta,\phi)$ to be dependent on the spherical harmonics, $Y_{lm}(\theta,\phi)$, in the following way:
\begin{align}
  Y^+_{lm}(\theta,\phi) = \mathcal{O}^{(1)}_{+}Y_{lm}(\theta,\phi) \nonumber\\
  Y^-_{lm}(\theta,\phi) = \mathcal{O}^{(1)}_{-}Y_{lm}(\theta,\phi)
\end{align}

This newfound knowledge allows us to remove all $\theta$ and $\phi$ dependence from the eigenvalue equations~\ref{eq:feigenproblems}, leaving us with
\begin{align}
   \omega \tilde{\Theta}_1 = \mathcal{H}^{(f)}_1 \tilde{\Theta}_1 \\
   \omega \tilde{\Theta}_2 = \mathcal{H}^{(f)}_2 \tilde{\Theta}_2
\end{align}

\noindent where the fermionic Hamiltonians $\mathcal{H}^{(f)}_i$ now take the form
{\footnotesize \begin{align}
   \mathcal{H}^{(f)}_1 &= \left(\begin{array}{l l l l l l l l}
                          -p & c_{3+}l(l+1) & c_{2+} & 0 & 0 & i c_+ & 0 & 0 \\
                          -c_{4-} & p & 0 & c_{1-} & 0 & 0 & 0 & 0 \\
                          -c_{1-} & 0 & p & -c_{3-}l(l+1) & 0 & 0 & 0 & i c_- \\
                          0 & -c_{2+} & c_{4+} & -p & 0 & 0 & 0 & 0 \\
                          0 & 0 & 0 & 0 & -p & -c_{3+} & -c_{2+} & 0 \\
                          i c_- & 0 & 0 & 0 & c_{4-}l(l+1) & p & 0 & -c_{1-} \\
                          0 & 0 & 0 & 0 & c_{1-} & 0 & p & c_{3-} \\
                          0 & 0 & i c_+ & 0 & 0 & c_{2+} & -c_{4+}l(l+1) & -p
                          \end{array}
                          \right)
\end{align}}

\noindent and
{\footnotesize \begin{align}
   \mathcal{H}^{(f)}_2 &= \left(\begin{array}{l l l l l l l l}
                          -p & -c_{3+} & c_{2+} & 0 & 0 & 0 & 0 & 0 \\
                          c_{4-}l(l+1) & p & 0 & c_{1-} & -i c_- & 0 & 0 & 0 \\
                          -c_{1-} & 0 & p & c_{3-} & 0 & 0 & 0 & 0 \\
                          0 & -c_{2+} & -c_{4+}l(l+1) & -p & 0 & 0 & -i c_+ & 0 \\
                          0 & -i c_+ & 0 & 0 & -p & c_{3+}l(l+1) & -c_{2+} & 0 \\
                          0 & 0 & 0 & 0 & -c_{4-} & p & 0 & -c_{1-} \\
                          0 & 0 & 0 & -i c_- & c_{1-} & 0 & p & -c_{3-}l(l+1) \\
                          0 & 0 & 0 & 0 & 0 & c_{2+} & c_{4+} & -p
                          \end{array}
                          \right)
\end{align}}

These two matrices have the same eight eigenvalues
\begin{align}
   \omega &= \left\{\begin{array}{l}
                    \pm\sqrt{c_{10}(p,l) + \sqrt{c_8(p,l)} \pm \sqrt{c_{9+}}(p,l)} \\
                    \pm\sqrt{c_{10}(p,l) - \sqrt{c_8(p,l)} \pm \sqrt{c_{9-}}(p,l)}
                    \end{array}
                    \right.
\end{align}

\noindent where
\begin{align}
  c_5 &= c_{12}^2 - 3c_{11}c_{13} + 12 c_{14} \nonumber\\
  c_6 &= 2 c_{12}^3 - 9 c_{12}(c_{11}c_{13} + 8 c_{14}) + 27(c_{13}^2 + c_{11}^2c_{14}) \nonumber\\
  c_7 &= c_6 + \sqrt{-4 c_5^3 + c_6^2} \nonumber\\
  c_8 &= c_{11}^2 + \frac{2}{3}\left(-4 c_{12} + \frac{2^{4/3} c_5}{c_7^{1/3}} + 2^{2/3} c_7^{1/3}\right) \nonumber\\
  c_{9\pm} &= \frac{2}{3}\left(3 c_{11}^2 - 8c_{12} - \frac{2^{4/3} c_5}{c_7^{1/3}} - 2^{2/3} c_{7}^{1/3} \pm \frac{3(4 c_{11}c_{12} - c_{11}^3- 8 c_{13})}{\sqrt{c_8}} \right)\nonumber\\
  c_{10} &= 2(p^2 - (c_{3+}c_{4-} + c_{3-}c_{4+})l(l+1) - c_{-}c_{+} - 2 c_{1-} c_{2+})
\end{align}

\noindent and where
\begin{align}
  c_{11} &= 4c_{1-} c_{2+} + 2 c_{-}c_{+} + (2 c_{3+} c_{4-} + 2 c_{3-} c_{4+}) l(l+1) - 4p^2 \nonumber\\
  c_{12} &= 6 c_{1-}^2c_{2+}^2 + c_{2+}^2 c_{-}^2 + 4 c_{1-}c_{2+} c_{-}c_{+} + c_{1-}^2 c_{+}^2 + c_{-}^2 c_{+}^2 +\nonumber\\
  &+ (2 c_{2+}^2 c_{3-} c_{4-} + 4 c_{1-} c_{2+} c_{3+} c_{4-} + 4 c_{1-} c_{2+} c_{3-} c_{4+} + 2 c_{1-}^2 c_{3+} c_{4+} + 2 c_{3+} c_{4-} c_{-} c_{+} + \nonumber\\
  &+  2 c_{3-} c_{4+} c_{-} c_{+})l(l+1) + (c_{3+}^2 c_{4-}^2 + 4 c_{3-} c_{3+} c_{4-} c_{4+} + c_{3-}^2 c_{4+}^2) l^2(l+1)^2 + (-12 c_{1-} c_{2+} + \nonumber\\
  &- 6 c_{-} c_{+} + (-6 c_{3+} c_{4-} - 6 c_{3-} c_{4+}) l(l+1)) p^2 + 6 p^4 \nonumber\\
c_{13} &= 4 c_{1-}^3 c_{2+}^3 + 2 c_{1-} c_{2+}^3 c_{-}^2 + 2 c_{1-}^2 c_{2+}^2 c_{-} c_{+} + 2 c_{1-}^3 c_{2+}c_{+}^2 + \nonumber\\
 &+2 c_{1-} c_{2+} c_{-}^2 c_{+}^2 + (4 c_{1-} c_{2+}^3 c_{3-} c_{4-} + 2 c_{1-}^2 c_{2+}^2 c_{3+} c_{4-} + \nonumber\\
    &+2 c_{1-}^2 c_{2+}^2 c_{3-} c_{4+} + 4 c_{1-}^3 c_{2+} c_{3+} c_{4+} + 2 c_{1-} c_{2+} c_{3+} c_{4+} c_{-}^2 + 2 c_{1-} c_{2+} c_{3+} c_{4-} c_{-} c_{+} + \nonumber\\
    &+ 2 c_{1+} c_{2+} c_{3-} c_{4+} c_{-} c_{+} + 2 c_{1-} c_{2+} c_{3-} c_{4-} c_{+}^2) l^2 (1 + l)^2 + \nonumber\\
    &+ (2 c_{2+}^2 c_{3-} c_{3+} c_{4-}^2 + 2 c_{2+}^2 c_{3-}^2 c_{4-} c_{4+} + 4 c_{1-} c_{2+} c_{3-} c_{3+} c_{4-} c_{4+} + 2 c_{1-}^2 c_{3+}^2 c_{4-} c_{4+} + \nonumber\\
    &+ 2 c_{1-}^2 c_{3-} c_{3+} c_{4+}^2 + c_{3+}^2 c_{4-}^2 c_{-} c_{+} + c_{3-}^2 c_{4+}^2 c_{-} c_{+}) l^4 (1 + l)^4 + (2 c_{3-} c_{3+}^2 c_{4-}^2 c_{4+} +  \nonumber\\
    & + 2 c_{3+}^2 c_{3+} c_{4-} c_{4+}^2) l^6 (1 + l)^6 + (-12 c_{1-}^2 c_{2+}^2 - 2 c_{2+}^2 c_{-}^2 - 8 c_{1-} c_{2+} c_{-} c_{+} - 2 c_{1-}^2 c_{+}^2 +\nonumber\\
    &- 2 c_{-}^2 c_{+}^2 + (-4 c_{2+}^2 c_{3-} c_{4-} - 8 c_{1-} c_{2+} c_{3+} c_{4-} - 8 c_{1-} c_{2+} c_{3-} c_{4+} - 4 c_{1-}^2 c_{3+} c_{4+} - 4 c_{3+} c_{4-} c_{-} c_{+} \nonumber\\
    &- 4 c_{3-} c_{4+} c_{-} c_{+}) l^2 (1 + l)^2 + (-2 c_{3+}^2 c_{4-}^2 - 8 c_{3-} c_{3+} c_{4-} c_{4+} - 2 c_{3-}^2 c_{4+}^2) l^4 (1 + l)^4) p^2 + (12 c_{1-} c_{2+} + \nonumber\\
    &+ 6 c_{-} c_{+} + (6 c_{3+} c_{4-} + 6 c_{3-} c_{4+}) l^2 (1 + l)^2) p^4 - 4 p^6 \nonumber\\
c_{14} &= c_{1-}^4 c_{2+}^4 + c_{1-}^2 c_{2+}^4 c_{-}^2 + c_{1-}^4 c_{2+}^2 c_{+}^2 + c_{1-}^2 c_{2+}^2 c_{-}^2 c_{+}^2 + (2 c_{1-}^2 c_{2+}^4 c_{3+} c_{4+} + \nonumber\\
    &+2 c_{1-}^4 c_{2+}^2 c_{3+} c_{4+} + 2 c_{1-}^2 c_{2+}^2 c_{3+} c_{4+} c_{-}^2 + 2 c_{1-}^2 c_{2+}^2 c_{3-} c_{4-} c_{+}^2) l^2 (1 + l)^2 + (c_{2+}^4 c_{3-}^2 c_{4-}^2 + \nonumber\\
    &+4 c_{1-}^2 c_{2+}^2 c_{3-} c_{3+} c_{4-} c_{4+} + c_{1-}^4 c_{3+}^2 c_{4+}^2 + c_{1-}^2 c_{3+}^2 c_{4+}^2 c_{-}^2 + c_{2+}^2 c_{3-}^2 c_{4-}^2 c_{+}^2) l^4 (1 + l)^4 + \nonumber\\
    &+(2 c_{2+}^2 c_{3-}^2 c_{3+} c_{4-}^2 c_{4+} + 2 c_{1-}^2 c_{3-} c_{3+}^2 c_{4-} c_{4+}^2) l^6 (1 + l)^6 +
 c_{3-}^2 c_{3+}^2 c_{4-}^2 c_{4+}^2 l^8 (1 + l)^8 + \nonumber\\
    &+(-4 c_{1-}^3 c_{2+}^3 - 2 c_{1-} c_{2+}^3 c_{-}^2 - 2 c_{1-}^2 c_{2+}^2 c_{-} c_{+} - 2 c_{1-}^3 c_{2+} c_{+}^2 - 2 c_{1-} c_{2+} c_{-}^2 c_{+}^2 + \nonumber\\
    &+ (-4 c_{1-} c_{2+}^3 c_{3-} c_{4-} - 2 c_{1-}^2 c_{2+}^2 c_{3+} c_{4-} - 2 c_{1-}^2 c_{2+}^2 c_{3-} c_{4+} - 4 c_{1-}^3 c_{2+} c_{3+} c_{4+} +\nonumber\\
    &- 2 c_{1-} c_{2+} c_{3+} c_{4+} c_{-}^2 - c_{1-} c_{2+} c_{3+} c_{4+} c_{-} c_{+} - 2 c_{1-} c_{2+} c_{3-} c_{4+} c_{-} c_{+} - 2 c_{1-} c_{2+} c_{3-} c_{4-} c_{+}^2) l^2 (1 + l)^2 + \nonumber\\
    &+(-2 c_{2+}^2 c_{3-} c_{3+} c_{4-}^2 - 2 c_{2+}^2 c_{3-}^2 c_{4-} c_{4+} - 4 c_{1-} c_{2+} c_{3-} c_{3+} c_{4-} c_{4+} - 2 c_{1-}^2 c_{3+}^2 c_{4-} c_{4+} +\nonumber\\
    &- 2 c_{1-}^2 c_{3-} c_{3+} c_{4+}^2 - c_{3+}^2 c_{4-}^2 c_{-}c_{+} - c_{3-}^2 c_{4+}^2 c_{-} c_{+}) l^4 (1 + l)^4 + (-2 c_{3-} c_{3+}^2 c_{4-}^2 c_{4+} +\nonumber\\
    &- 2 c_{3-}^2 c_{3+} c_{4-} c_{4+}^2) l^6 (1 + l)^6) p^2 + (6 c_{1-}^2 c_{2+}^2 + c_{2+}^2 c_{-}^2 + 4 c_{1-} c_{2+} c_{-} c_{+} + c_{1-}^2 c_{+}^2 + c_{-}^2 c_{+}^2 + \nonumber\\
    &+ (2 c_{2+}^2 c_{3-} c_{4-} + 4 c_{1-} c_{2+} c_{3+} c_{4-} + 4 c_{1-} c_{2+} c_{3-} c_{4+} + 2 c_{1-}^2 c_{3+} c_{4+} + 2 c_{3+} c_{4-} c_{-} c_{+} + \nonumber\\
    &+ 2 c_{3-} c_{4+} c_{-} c_{+}) l^2 (1 + l)^2 + (c_{3+}^2 c_{4-}^2 + 4 c_{3-} c_{3+} c_{4-} c_{4+} + c_{3-}^2 c_{4+}^2) l^4 (1 + l)^4) p^4 + \nonumber\\
    &+(-4 c_{1-} c_{2+} - 2 c_{-} c_{+} + (-2 c_{3+} c_{4-} - 2 c_{3-} c_{4+}) l^2 (1 + l)^2) p^6 + p^8
\end{align}

\section{$\zeta$-function Regularization of the Bosonic Fluctuation Energy}\label{app:regularization}
The four terms in the one loop bosonic energy
\begin{align}\label{eq:bosonicenergy}
  E_1^{(b)} =  \sum_{p}\omega_1 + 3\sum_{p,l,m}\omega_2 + \sum_{p,l,m}\omega_+ + \sum_{p,l,m} \omega_-
\end{align}

\noindent can all be written in the form
\begin{align}\label{eq:bosonicsumgeneral}
  \sum_{p,l,m} \omega = \sqrt{p^2 + f(l)},
\end{align}

\noindent where $f(l)$ is a function of only $l$.
Demanding vanishing of the bosonic eigenfunctions at $x=0,L$, leaves us with $p$ quantized to $p=n\pi/L$.  In addition we can easily perform the $m$ summation in~\ref{eq:bosonicsumgeneral}, and split up the $l$ summation up as so:
\begin{align}\label{eq:renormalizegeneral}
  \sum \omega &= \sum_{n,l,m} \sqrt{(n\pi/L)^2 + f(l)} \nonumber\\
              &= \sum_{n=1}^{\infty} \sqrt{(n\pi/L)^2 + f(0)} + \sum_{n=1,l=1}^{\infty}(2l+1)\sqrt{(n\pi/L)^2 + f(l)}
\end{align}

The function $f(l)$ for the first sum in equation~\ref{eq:bosonicenergy} is actually $l$-independent
\begin{align}
  f(l) &= f_1 = g_{xx}m_{\Psi}^2
\end{align}

\noindent We have for this sum then only the leftmost term from equation~\ref{eq:renormalizegeneral}
\begin{align}\label{eq:1sum}
  \sum_p \omega_1 &= \sum_{n=1}^{\infty} \sqrt{(n\pi/L)^2 + f_1} \nonumber\\
                &= \sqrt{f_1} \sum_{n=1}^{\infty} \sqrt{1 + \frac{n^2\pi^2}{L^2f_1}} \nonumber\\
                &= \sqrt{f_1} \sum_{n=1}^{\infty} \sum_{q=0}^{\infty} \binom{\frac{1}{2}}{q} \left(\frac{n^2\pi^2}{L^2f_1}\right)^q \nonumber\\
                &= \sqrt{f_1}  \sum_{q=0}^{\infty} \binom{\frac{1}{2}}{q} \left(\frac{\pi^2}{L^2 f_{1}}\right)^q\sum_{n=1}^{\infty}n^{2q} \nonumber\\
                &= \sqrt{f_1}  \sum_{q=0}^{\infty} \binom{\frac{1}{2}}{q} \left(\frac{\pi^2}{L^2 f_{1}}\right)^q\zeta(0) ?\delta_q^0? \nonumber\\
                &= \sqrt{f_1} \zeta(0) = -\frac{1}{2}\sqrt{g_{xx}}m_{\Psi}
\end{align}

\noindent Notice we find no Luscher term associated with this oscillation, merely a constant energy contribution.  We will take the common point of view that constants such as this do not actually contribute to the ground state energy, focusing merely on Luscher terms, i.e., terms proportional to $1/L$.

We look now to the next sum in equation~\ref{eq:bosonicenergy}, which has $l$ dependence given by
\begin{align}
  f(l) = f_2(l) &= g_{xx} \frac{R}{2}l(l+1).
\end{align}

\noindent Which means we will need to analyze both terms in~\ref{eq:renormalizegeneral}
\begin{align}\label{eq:2sum}
  \sum \omega_2 &= \sum_{n=1}^{\infty} n\pi/L + \sum_{n=1,l=1}^{\infty}(2l+1)\sqrt{(n\pi/L)^2 + f_2(l)} \nonumber\\
                  &= \frac{\pi}{L}\sum_{n=1}^{\infty}n + \sum_{n=1,l=1}(2l+1)\sqrt{f_2(l)}\sqrt{1 + \frac{n^2\pi^2}{L^2f_2(l)}} \nonumber\\
                  &= \frac{\pi}{L}\zeta(-1) + \sum_{l=1}(2l+1)\sqrt{f_2(l)}\sum_{q=0}^{\infty}\binom{\frac{1}{2}}{q} \left(\frac{\pi^2}{L^2f_2(l)}\right)^q\sum_{n=1}^{\infty}n^{2q}\nonumber\\
                  &=\frac{\pi}{L}\zeta(-1) + \sum_{l=1}(2l+1)\sqrt{f_2(l)}\sum_{q=0}^{\infty}\binom{\frac{1}{2}}{q} \left(\frac{\pi^2}{L^2f_2(l)}\right)^q \zeta(0)?\delta_q^0?\nonumber\\
                  &= \frac{\pi}{L}\zeta(-1) + \sum_{l=1}(2l+1)\sqrt{g_{xx}\frac{R}{2}l(l+1)}\zeta(0) \nonumber\\
                  &= \frac{\pi}{L}\zeta(-1) + \zeta(0)\sqrt{g_{xx}\frac{R}{2}}f_{\zeta}\left(\frac{1}{2}\right) 
\end{align}

\noindent The function $f_{\zeta}(s)$ is defined as
\begin{align}
  f_{\zeta}(s) \equiv \sum_{l=1}^{\infty}(2l+1)\sqrt{l(l+1)}.
\end{align}

\noindent Through $\zeta$-function regularization, we find $f_{\zeta}(\frac{1}{2}) \approx -0.265096$.
Here we find a Luscher term, $\frac{\pi}{L}\zeta(-1)$, contribution to the zero point energy.

Finally, we regularize the final two sums in equation~\ref{eq:bosonicenergy}, whose $l$ dependence can be written succinctly for both terms as
\begin{align}
  f(l) =  f_{\pm}(l) &= a_1 l(l+1) + a_2 \pm a_2\sqrt{1 + a_3 l(l+1)}
\end{align}

\noindent with
\begin{align}
  a_1 &= g_{xx}\frac{R}{2},~~~a_2=g_{xx}\frac{m_{\tau}^2}{2},~~~a_3=\frac{16\pi T_0 R}{9 b^3 g_s M m_{\tau}^4}
\end{align}

Through a series of binomial expansions and a $\zeta$-function regularization, we calculate the final two sums to be
\begin{align}
  \sum \omega_{\pm} = \sum_{n=1}^{\infty} \sqrt{\left(\frac{n\pi}{L}\right)^2 + f_{\pm}(0)} + a_4
\end{align}

\noindent where
\begin{align}
  a_4 &\equiv \zeta(0) \sqrt{a_2} \sum_{q=0}^{\infty} \sum_{r=0}^{r<q}
  \sum_{v=0}^{v=\left\{ \stackrel{\infty, \mbox{{\tiny odd~r}}}{\frac{r}{2}, \mbox{{\tiny even~r}}} \right.}
  \binom{\frac{1}{2}}{q}\binom{q}{r}\binom{\frac{r}{2}}{v} (\pm 1)^r \left(\frac{a_1}{a_2}\right)^{q-r} a_3^v
  f_{\zeta}(q-r+v)
\end{align}

\noindent and
\begin{align}
  f_+(0) = 2 a_2 = g_{xx}m_{\tau}^2,~~~~f_-(0) = 0.
\end{align}

It is easy to show that
\begin{align}\label{eq:pmsum}
  \sum \omega_+ &= \zeta(0)\sqrt{f_+(0)} + a_4 = -\frac{1}{2} \sqrt{g_{xx}}m_{\tau} + a_4, \nonumber\\
  \sum \omega_- &= \frac{\pi}{L}\zeta(-1) + a_4. 
\end{align}

Using equations~\ref{eq:1sum},~\ref{eq:2sum},and~\ref{eq:pmsum} with equation~\ref{eq:bosonicenergy}, we calculate the one loop bosonic energy to be
\begin{align}\label{eq:E1b}
  E_1^{(b)} &= V_{\mbox{\small{L\"uscher}}} + V_c(k) \\
  V_{\mbox{\small{L\"uscher}}} &= 4 \frac{\pi}{L}\zeta(-1) = -\frac{\pi}{3L} \\
  V_c(k) &= - \frac{1}{2} \sqrt{g_{xx}}(m_{\tau} + m_{\Psi}) + 3 \zeta(0)\sqrt{g_{xx}\frac{R}{2}} f_{\zeta}\left(\frac{1}{2}\right) + 2 a_4.
\end{align}

\noindent The first term, $V_{\mbox{\small{L\"uscher}}}$, is a L\"uscher term, signified by the $1/L$ dependence.  The second term, $V_c(k)$, is independent of $L$, the \emph{length} of the $k$-string, and is merely an additive constant.  The Luscher term, obviously respects the $k \to M - k$ symmetry, as it is independent of $k$.  This symmetry, however, is not at all obvious for $V_c(k)$.  Through careful inspection of $V_c(k)$, we find that it's $k$ dependence depends on only the following list:
\begin{align}
 &g_{xx},~m_{\tau},~m_{\Psi},~\mbox{and}~R.
\end{align}

\begin{figure}
  \centering
  \includegraphics[width=0.45\columnwidth]{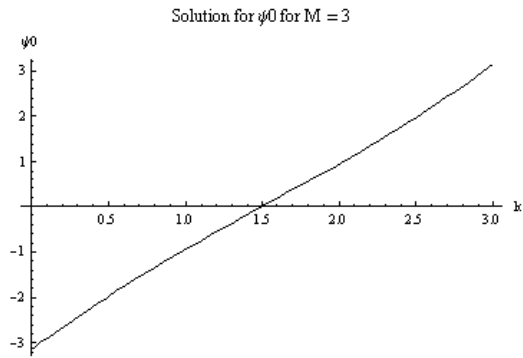}
  \caption{Solution of transcendental equation~\ref{eq:psicondition} for $M = 3$.  Notice that the solution is anti-symmetric under the transformation $k \to M - k$.  This means that the additive constant to the bosonic energy in equation~\ref{eq:E1b} is invariant under this transformation.  This holds true for any value of $M$.}
  \label{fig:Psi0solution}
\end{figure}

Through their definitions in equations~\ref{eq:constants} and~\ref{eq:Psimass}, we discover that their $k$-dependence lies only in $\cos\psi_0$.  So if the transcendental solution for $\psi_0$,
\begin{align}\tag{\ref{eq:psicondition}}
   \psi_0 - \frac{2k\pi}{M} + \pi &= (b^2 - 1)\sin\psi_0 \nonumber,
\end{align}

\noindent respects the symmetry $\psi_0 \to \pm \psi_0$ under exchange of $k$ quarks with $k$ anti-quarks ($k \to M - k$), then $\cos\psi_0$ will remain invariant, and thus, so will $V_c(k)$.  We find that the transcendental solution \emph{exactly} satisfies this condition, as shown in figure~\ref{fig:Psi0solution}.

\eject


\end{document}